\documentclass[12pt,preprint]{aastex}

\def\farc{\hbox{$\ \!\!^{\prime\prime}$}}
\def\fs{\hbox{$.\!\!^{\rm s}$}}

\slugcomment{For submission to the Astrophysical Journal}

\shorttitle{Molecular Gas Deposited by X-ray Cooling Flow}
\shortauthors{Lim et al.}

\begin{document}


\title{Radially-Inflowing Molecular Gas in NGC~1275 \\ Deposited by a X-ray Cooling Flow in the Perseus Cluster}


\author{Jeremy Lim}
\affil{Institute of Astronomy \& Astrophysics, Academia Sinica, PO Box 23-141, Taipei 10617, Taiwan}
\email{jlim@asiaa.sinica.edu.tw}

\author{YiPing Ao\altaffilmark{1}}
\affil{Purple Mountain Observatory, Chinese Academy of Sciences, Nanjing 210008, \\ People's Republic of China}
\email{ypao@pmo.ac.cn}

\and

\author{Dinh-V-Trung}
\affil{Institute of Astronomy \& Astrophysics, Academia Sinica, PO Box 23-141, Taipei 10617, Taiwan}
\email{trung@asiaa.sinica.edu.tw}


\altaffiltext{1}{Visiting Scholar, Institute of Astronomy \& Astrophysics, Academia Sinica}


\begin{abstract}

We have imaged in CO(2-1) the molecular gas in NGC~1275 (Perseus~A), the cD galaxy at the center of the Perseus Cluster, at a spatial resolution of $\sim$1~kpc over a central region of radius $\sim$10 kpc.   Per~A is known to contain $\sim$$1.3 \times 10^{10} {\rm \ M_{\odot}}$ of molecular gas, which has been proposed to be captured from mergers with or ram-pressure stripping of gas-rich galaxies, or accreted from a X-ray cooling flow.  The molecular gas detected in our image has a total mass of $\sim$$4 \times 10^{9} {\rm \ M_{\odot}}$, and for the first time can be seen to be concentrated in three radial filaments with lengths ranging from at least 1.1--2.4~kpc all lying in the east-west directions spanning the center of the galaxy to radii of $\sim$8~kpc.  The eastern and outer western filaments exhibit larger blueshifted velocities with decreasing radii, whereas the inner western filament spans the systemic velocity of the galaxy.  The molecular gas shows no signature of orbital motion, and is therefore unlikely to have been captured from gas-rich galaxies.  Instead, we are able to reproduce the observed kinematics of the two outer filaments as free-fall in the gravitational potential of Per~A, as would be expected if they originate from a X-ray cooling flow.  Indeed, all three filaments lie between two prominent X-ray cavities carved out by radio jets from Per~A, and closely resembles the spatial distribution of the coolest X-ray gas in the cluster core.  The inferred mass-deposition rate into the two outermost filaments alone is roughly $75 {\rm \ M_{\odot} \ yr^{-1}}$.  This cooling flow can provide a nearly continuous supply of molecular gas to fuel the active nucleus in Per~A.

\end{abstract}


\section{Introduction}
Rich groups and clusters of galaxies are immersed in hot ($\sim$$10^7$--$10^8$~K) X-ray-emitting gas that constitutes a large fraction of their baryonic mass \citep{sar86}.  Radiative cooling of this gas, by virtue of its X-ray emission, should result in an inflow of relatively cool gas to the cluster center; i.e., a X-ray cooling flow \citep{cow77,fab77}.  In the absence of any reheating, this flow is inferred to have mass-deposition rates ranging from $\sim$$10 {\rm \ M_\sun \ yr^{-1}}$ to $\sim$$1000 {\rm \ M_\sun \ yr^{-1}}$ depending on the given cluster, and over a Hubble time ($\sim$$10^{10} {\rm \ yr}$) should deposit $\sim$$10^{11} {\rm \ M_\sun}$ to $\sim$$10^{13} {\rm \ M_\sun}$ of relatively cool gas in the central cD (giant elliptical) galaxy \citep[e.g., reviews by][]{fab94,all00}.  

Numerous searches have been made for the predicted mass sink from X-ray cooling flows.  These searches cover a broad range of temperatures, from ionized gas at temperatures of $\sim$$10^{5.5} {\rm \ K}$ \citep[e.g., through the OVI line in the ultraviolet;][and references therein]{bre06} down to $\sim$$10^4 {\rm \ K}$ \citep[e.g., H$\rm \alpha$, $\rm {[NII]}$ and $\rm {[SII]}$ lines in the optical;][and references therein]{cra92}, to neutral gas at temperatures $\lesssim 10^3 {\rm \ K}$ in atomic hydrogen \citep[21-cm line;][and references therein]{ode98}, hot molecular hydrogen \citep[near-IR vibrational lines;][and references therein]{edg02}, as well as cool molecular hydrogen \citep[traced using rotational lines of carbon monoxide at mm wavelengths;][and references therein]{edg01,sal03}.  None of these searches have found gas at quantities anywhere near the abovementioned predicted levels (prior to the era of the XMM-Newton satellite as explained below) in putative cooling-flow clusters \citep[e.g., review by][]{fab94,mat03}.

Over the past few years, observations of hot gas in galaxy clusters with the XMM-Newton and Chandra X-ray observatories have radically altered our concept of X-ray cooling flows.  Spectra taken with XMM-Newton at high energy resolution and sensitivity indeed show a decrease in gas temperature towards the cluster center, but fail to reveal any detectable gas at temperatures below about one-third the bulk ambient temperature \citep{pet03,pet06}.  These results provide no evidence for a cooling flow, and imply that any gas cooling to temperatures $\lesssim 1$--$3 \times 10^7 {\rm \ K}$ has a mass-deposition rate at least ten times lower than that previously inferred.  Furthermore, images taken with Chandra at high angular resolutions reveal that the X-ray gas at the centers of putative cooling-flow clusters is usually disturbed \cite[e.g.,][]{fab00,mcn00,bla01,mcn01,maz02}.  X-ray cavities are commonly seen, spatially coincident with radio jets from the central galaxy where detectable \citep[][and references therein]{dun05}, indicating that energetic particles injected from an active galactic nucleus (AGN) are responsible for the observed disturbances.  The work needed to inflate these X-ray cavities implies that the energy output of AGNs is more than sufficient to balance radiative losses from the X-ray gas, and can therefore greatly diminish if not quench the cooling flow \citep{chu02,bir04}.

Against this backdrop, in the last few years searches targeting putative strong cooling-flow clusters have become relatively successful at detecting massive amounts of cool molecular hydrogen gas as traced in carbon monoxide (CO) at the centers of these clusters \citep{edg01,sal03}.  About twenty such examples are now known, whereas earlier searches \citep[e.g.,][]{gra90,mcn94,ant94,bra94,ode94} detected CO gas at the center of only one cluster, the Perseus Cluster \citep{laz89,mir89}.  The earlier searches targeted relatively nearby clusters which, in the light of the downward revisions in mass-deposition rates, are now known to mostly have very weak if any cooling flows.  On the other hand, the more recent surveys \citep{edg01,sal03} target more distant clusters with higher mass-deposition rates as inferred in the traditional picture of X-ray cooling flows.  Follow-up imaging where performed reveals that the molecular gas is usually located within a region of $< 20$~kpc coincident with the central galaxy \citep{edg03,sal04}.  The inferred mass of molecular hydrogen gas spans the range $\sim$$10^9$--$10^{11} {\rm \ M_\sun}$, still much lower than that predicted in the traditional picture of X-ray cooling flows, but closer now to the revised picture given the current upper limits in mass-deposition rates.  The amount of cool molecular gas far outweighs all other gas components detected at temperatures below $\sim$$10^6 {\rm \ K}$ combined.  These results leave open the possibility that X-ray cooling flows may deposit what is still a massive amount of cool gas compared with that normally present in galaxies (especially elliptical or cD galaxies).

A cooling flow, however, is not the only possible source of molecular or other relatively cool gas in cD galaxies.  Ram-pressure stripping of the interstellar medium of gas-rich galaxies by the hot intracluster medium, or cannibalisms of gas-rich galaxies by the cD galaxy, provide plausible if not more attractive alternatives \citep[e.g., reviews by][]{mat03,bre04}.  Unlike cooling flows, ram-pressure stripping is firmly established to occur in cluster galaxies \citep[e.g., review by][]{bos06}, and likewise mergers and cannibalisms in field galaxies \citep[e.g., review by][]{bar92}.  Indeed, cD galaxies presumably attained their present-day mass through repeated cannibalisms of cluster galaxies, or repeated merging of subclusters and galaxy groups before or during cluster virialization \citep[e.g., see discussion in][]{pim06}.  In support of these arguments, a number of elliptical galaxies in the field, members of groups not recognized to have X-ray cooling flows, or those residing in but not at the centers of clusters have been detected in CO with inferred molecular hydrogen gas masses of up to several $\sim$$10^9 {\rm \ M_\sun}$ \citep[see][and references therein]{kna96,lim00,lim04}.  At least one of these galaxies, NGC~759, resides in but not at the center of a putative cooling-flow cluster, Abell~262.  NGC~759 has an inferred mass in molecular hydrogen gas of $2.4 \times 10^9 {\rm \ M_\sun}$, at least two orders of magnitude higher than the upper limit placed on the mass of molecular hydrogen gas in the central cD galaxy, NGC~708 \citep{bra94}.  These results clearly demonstrate that a cooling flow is not the only mechanism capable of depositing large quantities of molecular gas in cluster elliptical galaxies.

At the present time, the available observations do not allow us to confidently discriminate between the different possible origins for the molecular (or other relatively cool) gas in cD galaxies.  Yet, naively, such a distinction should be possible simply from the observed spatial distribution and kinematics of the molecular gas.  In the absence of any disturbances, the molecular gas deposited by a X-ray cooling flow should simply flow radially inwards along the gravitational potential of the cD galaxy.  On the other hand, gas captured from a gas-rich galaxy is expected to orbit (and hence exhibit a global rotation) about the cD galaxy, reflecting the orbit (or passage) of the gas-rich galaxy around (or through) the cluster center or cD galaxy.  Here, we present a high angular-resolution CO observation of the cD galaxy Perseus~A (Per~A; NGC 1275) that reveals for the first time an ordered spatial-kinematic pattern in its molecular gas.  Located at a comparatively close distance of $\sim$74~Mpc (redshift $z=0.01756$, and assuming $H_0 = 70 {\rm \ km \ s^{-1} \ Mpc^{-1}}$ and $\Omega=1$), Per~A lies at the center of the Perseus Cluster (Abell~426), the X-ray-brightest cluster in the sky.  The radiative cooling time in the inner few tens of kpc of the Perseus cluster is less than $\sim$$10^9 {\rm \ yr}$, leading to a mass-deposition rate of a few $100 {\rm \ M_\sun \ yr^{-1}}$ in the absence of any heat input \citep{fab94,fab06,dun06}, the highest of any cluster within its distance.

For over a decade, Perseus~A was the only cD galaxy in a putative cooling-flow cluster known to contain molecular gas, which together with its other peculiarities and relative proximity have made it the subject of intensive studies.  This galaxy has been mapped with single-dish telescopes in the $J \rightarrow 1-0$ \citep{reu93}, $J \rightarrow 2-1$ \citep{reu93,bri98,sal06}, and $J \rightarrow 3-2$ \citep{bri98} transitions of CO.  The most recent map in CO(2-1) by \citet{sal06}, which has the highest angular resolution and covers the widest area, shows that the molecular gas is elongated approximately east-west across the center of the galaxy.  In agreement with previous single-dish maps, the molecular gas shows no apparent ordered spatial-kinematic pattern, with predominantly blueshifted velocities both east and west of the nucleus, and velocities near the systemic velocity immediately around the nucleus.  The inferred mass of molecular hydrogen gas in Per~A is $\sim$$1.3 \times 10^{10} {\rm \ M_{\odot}}$ \citep[Fig.~4 of][]{sal06}, far larger than that yet detected in field or non-cD cluster elliptical galaxies in the Local Universe \citep[e.g., review by][]{kna99}.  The only previous inteferometric map of the molecular gas in Per~A was made in CO(1-0) with the Nobeyama Millimeter Array (NMA) by \citet{ino96}.  This map differs in a number of important respects to single-dish maps \citep[see discussion in][]{sal06}, and bears little resemblence to that shown here in CO(2-1).  We suspect that the CO(1-0) interferometric map may be compromised by inaccurate bandpass calibration, as explained in the next section.  The only other published attempt to image the molecular gas in Per~A with an interferometer was made by \citet{bra95} in CO(2-1) with the IRAM Plateau de Bure Interferometer (PdBI), where virtually all the emission was resolved out.

Readers interested in the technical details of our observations and data reduction should now proceed to $\S2$.  Those interested only in the results can skip ahead to $\S3$.  In $\S4$ we compare the observed properties of the molecular gas with predictions for capture from gas-rich galaxies, and in $\S5$ with predictions for accretion from a X-ray cooling flow.  In $\S6$, we discuss the implications of our results for the cooling flow in the Perseus Cluster.  Readers interested only in a concise summary of our results and interpretation can proceed directly to $\S7$.

\section{Observations and Data Reduction}
Mapping of line emission from Per~A at radio wavelengths is made difficult by the much stronger continuum emission of the galaxy.  Extended jets and lobes dominate the continuum at centimeter wavelengths \citep{ped90}, whereas a compact nuclear source dominates the continuum at millimeter wavelengths \citep{bra95,ino96}.  In such a situation, accurate measurements of the spectral response (passband) of each antenna and its variation with time are required.  Any measurement errors introduce a frequency and time dependence in the point spread function of the telescope, which when convolved with the continuum source leave spatially-extended and frequency-dependent artifacts in the map that can be mistaken for or alter the line emission.  We suspect that such artifacts may be responsible for the differences seen between the interferometric map of \citet{ino96} in CO(1-0) (rest wavelength at 2.6~mm) and the single-dish maps mentioned in $\S1$, as well as the map made here.

The continuum emission of Per~A is weaker at shorter wavelengths, and hence the required measurement accuracy and stability of antenna passbands become correspondingly less stringent at higher transitions of CO.  We have therefore chosen to map Per~A in CO(2-1) at 1.3~mm, using the SubMillimeter Array (SMA).  The observations spanned $\sim$5.5 hrs with an on-source time of $\sim$4~hrs on Oct~22, and spanned $\sim$10~hrs with an on-source time of $\sim$7~hrs on Oct~24.  The weather was relatively poor on Oct~22, and so we report here only the observation on Oct~24. The SMA was in its compact configuration, and with 6 antennas operating provided projected baselines over the observation ranging from about 8--68~m (6--52~k$\lambda$).  We observed the CO(2-1) line in the upper sideband of the double-sideband receivers.  The correlator was configured to provide twenty-four independent spectral windows ("chunks") in each sideband.  Each window was divided into 128 channels with a channel width of 0.8125 MHz (i.e., $\sim$$1.1 {\rm \ km \ s^{-1}}$), giving a total frequency coverage in each sideband (with a small overlap between windows at their edges) of 1.97 GHz (i.e., $\sim$2590~km/s).

To accurately measure and track variations in the antenna passbands with time, we observed Mars at hourly intervals over the duration of our observation.  We found that the antenna passbands changed only slowly with time, and at a level smaller than that required to keep artifacts below the level of random noise
fluctuations.  We have therefore averaged all our measurements of Mars to derive a precise passband solution for each antenna.  Callisto served as the absolute flux calibrator.  We used the strong ($\sim$3.4~Jy) and compact (unresolved) nuclear continuum source in Per~A to derive the complex gains (amplitude and phase) of each antenna as a function of time, solving for the phase at 3~min intervals and amplitude at 20~min intervals.

The full-width half-maximum (FWHM) of the SMA primary beam at a frequency of 226.6~GHz in CO(2-1) is 55\farc, corresponding (with $1\farc = 360 {\rm \ pc}$ in the assumed Cosmology) to a circle of radius $\sim$10~kpc.  The CO(2-1) map of \citet{sal06} shows that the molecular gas extends to radii of $\sim$14~kpc both east and west of the nucleus, and hence our map spans only the inner central extent of the molecular gas.  Our observation was centered on the nucleus of Per~A, located at ${\rm Right \ Ascension = 03^h \ 19^m \ 48\fs16}$ and ${\rm Declination}=41^{\circ} \ 30\farc \ 42\farcs1$.  We attained a synthesized beam (with natural weighting of the visibilities) of $3\farcs0 \times 2\farcs7$ ($1.1 \times 1.0 {\rm \ kpc}$), with major axis at a position angle of $39^{\circ}.5$.  To make maps in CO~(2-1), we first derived the continuum emission from the line-free channels and subtracted this emission from the measured visibilities.  We then averaged the data in velocity channels of $20 {\rm \ km \ s^{-1}}$, and deconvolved the point spread function of the telescope from the image in each channel.  To make the corresponding maps in total intensity (zeroth moment) and intensity-weighted mean velocity (first moment), we first convolved the channel maps with a triangular function (i.e., Hanning smoothing) having a full-width at zero intensity of $100 {\rm \ km \ s^{-1}}$, and then discarded pixels with intensities less than $1.5 \sigma$ (where $\sigma$ is the root-mean-square noise fluctuation in one channel).  We then summed the remaining data in the channel maps (with no smoothing) to produce the total intensity map, and computed the mean velocity of each pixel weighted by its intensity in each channel to produce the intensity-weighted mean velocity map.

\section{Results}

\subsection{Spatial distribution and kinematics}
The CO(2-1) channel maps are shown in Figure~1, and the corresponding moment maps in Figure~2.  In Figure~1, the line intensity in each channel is plotted as contours on a grayscale map of the integrated line intensity.  In Figure~2, the integrated line intensity is plotted as contours on a color-coded map of the intensity-weighted mean velocity.  The location of the central continuum source, corresponding to the AGN, is indicated by a cross in both figures.  The velocities are measured with respect to the systemic heliocentric velocity of Per~A derived in the optical of $cz=5264 \pm 11 {\rm \ km \ s^{-1}}$ \citep{huc99}, where $c$ is the speed of light and $z$ the redshift.  For clarity, no primary beam correction has been applied to the channel maps shown in Figure~1.  On the other hand, to faithfully reproduce the overall intensity distribution of the molecular gas, a primary beam correction has been made to the integrated line intensity map shown in both Figure~1 and Figure~2. 

As in single-dish maps \citep{reu93,bri98,sal06}, the molecular gas in our interferometer map is concentrated primarily in the east-west directions.  Also as in single dish maps, the emission away from the nucleus on both the east and west, as well as north and south, is entirely blueshifted, with emission closest to the nucleus redshifted.  For the first time, our interferometer map reveals that the molecular gas in the central region of radius $\sim$10~kpc is concentrated primarily in three radial filaments, labeled in Figure~2 as the inner, western, and eastern filaments in order of increasing projected radii from the nucleus.  (As explained below, the inner filament may not actually comprise a single integral filament but rather two separate structures.)  Although the inner filament appears to connect with the western filament, they span very different velocities and have entirely different radial velocity gradients.  These filaments appear as distinct spatial-kinematic features in the channel maps (Fig.~1), and are clearly separate structures.  Similarly, the feature labeled E1 in Figure~2 is discontinuous in velocity from the eastern filament, and is regarded here as a separate structure (blob).  Another blob labeled E2 can be seen just to the north of the eastern filament.  An apparently isolated blob labeled W1 lies to the west of the western filament, located along a line from center that nearly passes through the inner and western filaments.  The only features located away from the east-west directions are the isolated blobs N1 and S1 located to the north and south of center respectively.  The spectra of all the features labeled in Figure~2 are shown in Figure~3.

Our CO(2-1) interferometer map bears little resemblance to the CO(1-0) interferometer map of \citet{ino96}, which as mentioned in $\S1$ also shows significant differences compared with single-dish maps.  First, none of the eastern features seen in Figure~2 are detected in the CO(1-0) interferometer map, despite the presence of gas at this location in single-dish maps.  Second, the inner and western filaments do not appear as distinct features in the CO(1-0) map, but coincide with a complex and spatially extended distribution of gas in the CO(1-0) map.  Third, gas in the CO(1-0) map at the location of the western filament exhibits both blueshifted and (primarily) redshifted velocities; in our map, the western filament exhibits only blueshifted velocities, consistent with that seen at this location in single-dish maps.  Finally, the CO(1-0) map shows gas towards the south-east of the nucleus that is not seen in our CO(2-1) map.  We conclude that the CO(1-0) interferometer map of \citet{ino96} is unreliable, and cannot be meaningfully compared with our CO(2-1) map.

The radial positions and extents, as well as velocities, of all the features labeled in Figure~2 are listed in Table~1.  To estimate the lengths of the filaments, we measure from the channel maps the difference in radial positions of the emission centroids at opposite ends in velocity spanned by that filament.  Because the emission is spatially resolved in many channels, this method provides a conservative estimate of the true length of the filaments, but is necessary in particular for the eastern and western filaments as they do not have a smoothly-varying intensity profile along their lengths.  As measured from the emission centroid in the channel maps, the inner filament can be traced from a (projected) radius of $1\farcs2 \pm 0\farcs3$ ($432 \pm 108$~pc) where it has a (projected) velocity of $100 {\rm \ km \ s^{-1}}$ to a radius of $4\farcs2 \pm 0\farcs3$ ($1.5 \pm 0.1$~kpc) westwards where it has a velocity of $-100 {\rm \ km \ s^{-1}}$, thus spanning a (projected) length of $\sim$3\farcs0 ($\sim$1.1~kpc).  A two-dimensional Gaussian fit to the inner filament in the integrated intensity map of Figure~2 gives a length (at full-width half maximum) of $4\farcs7 \pm 0\farcs6$ ($1.7 \pm 0.3$~kpc).  This filament straddles the systemic velocity of the galaxy, and exhibits a radial velocity gradient with smaller blueshifted velocities or larger redshifted velocities with decreasing radii (Fig.~2).  

The western filament can be traced from a radius of $9\farcs0 \pm 0\farcs3$ ($3.2 \pm 0.1$~kpc) where it has a velocity of $-180 {\rm \ km \ s^{-1}}$ to a radius of $15\farcs7 \pm 0\farcs4$ ($5.7 \pm 0.1$~kpc) westwards where it has a velocity of $-80 {\rm \ km \ s^{-1}}$, thus spanning a length of $\sim$6\farcs7 ($\sim$2.4~kpc).  The eastern filament can be traced from a radius of $15\farcs2 \pm 0\farcs4$ ($5.5 \pm 0.1$~kpc) where it has a velocity of $-160 {\rm \ km \ s^{-1}}$ to a radius of $21\farcs3 \pm 0\farcs3$ ($7.7 \pm 0.1$~kpc) eastwards where it has a velocity of $-40 {\rm \ km \ s^{-1}}$, thus spanning a length of 6\farcs1 (2.2~kpc).  Unlike the inner filament, the two outer filaments (and all the blobs) exhibit entirely blueshifted velocities, consistent with the velocities measured in single-dish maps at their locations.  These filaments can be seen to exhibit a radial velocity gradient with larger blueshifted velocities at smaller radii, opposite to that exhibited by the inner filament.

In Figure~4, we plot the CO(2-1) intensity of each filament along their radial extents as a function of velocity (i.e., position-velocity (PV-) diagrams).  For clarity, no primary beam correction has been applied to these plots.  Both the eastern and western filaments exhibit a simple linear change in velocity with radius from the center of the galaxy.  The feature at a position of 26\farc\ and velocity of $-80 {\rm \ km \ s^{-1}}$ in the PV-diagram of the eastern filament (bottom panel of Fig.~4) corresponds to the feature E1 (Fig.~2), which as mentioned earlier probably consitutes a separate structure.  The PV-diagram of the inner filament is more complicated: the velocity changes linearly with radius at both blueshifted and redshifted velocities, but with different velocity gradients on either sides of the systemic velocity.  It is possible that the redshift and blueshifted sides of the inner filament correspond to separate structures that are poorly resolved spatially in our map.  Indeed, a close examination of Figure~2 suggests that the redshifted and blueshifted portions of this filament lies at somewhat different position angles, producing a visible twist in the major axis of this filament between the two portions.

\subsection{Molecular gas mass}
The integrated line intensities, corrected for the primary beam response of the SMA, and corresponding molecular hydrogen (H$_2$) gas masses of all the features labeled in Figure~2 are listed in Table~1.  To facilitate a direct comparison with the molecular gas masses derived in \citet{sal06}, we adopt like them a brightness temperature in CO(2-1) of 0.7 times that in CO(1-0), and the standard CO(1-0) to H$_2$ conversion ratio (applicable to gas with solar metallicity) of $2.3 \times 10^{20} {\rm \ cm^{-2}}$ \citep{sol97}.  Altogether, the features have a total gas mass of $(4.2 \pm 0.2) \times 10^9 {\rm \ M_{\odot}}$, about half that detected in observations with single-dish telescopes over the same radius of $\sim$$1 \times 10^{10} {\rm \ M_{\odot}}$ \citep[Fig.~4 of][]{sal06}.  The three filaments alone have a total gas mass of $(3.3 \pm 0.2) \times 10^9 {\rm \ M_{\odot}}$, about $80\%$ of that contained in all the features.  Our interferometric observation is preferentially sensitive to relatively bright and compact features, and hence traces regions where the molecular gas is most strongly concentrated (i.e., highest filling factor).

To determine if there is emission external to the features detected in Figure~2 that lies below our brightness sensitivity, we have integrated the channel maps over the entire SMA primary beam from a velocity of $-230 {\rm \ km \ s^{-1}}$ to $170 {\rm \ km \ s^{-1}}$.  The integrated line intensity determined in this manner is $409 \pm 28 {\rm \ Jy \ km \ s^{-1}}$, corresponding to a molecular gas mass of $(11.8 \pm 0.8) \times 10^9 {\rm \ M_{\odot}}$.  This is higher than the total molecular gas mass of the features detected in Figure~2, and comparable with the abovementioned molecular gas mass detected with single-dish telescopes over the same region.

\subsection{Correspondence with H$\alpha$ gas}
Per~A exhibits a complex H$\alpha$ nebula that extends beyond the optically visible confines of the galaxy to nearly 70~kpc \citep{con01}.  The origin and excitation of this nebula is very poorly understood.  In Figure~5, we plot the integrated CO(2-1) intensity on a color map of the inner region of this H$\alpha$ nebula.  All the molecular filaments and blobs lying in the east-west directions, but not the blobs lying north and south of center, appear to be associated with the H$\alpha$ nebula.  The inner filament closely coincides with a bright western extension in H$\alpha$ from a compact nuclear component.  The inner radius of the western filament closely coincides with an apparent extension of the bright central H$\alpha$ feature further west.  The eastern filament coincides with local brightenings in H$\alpha$, as do the nearby blobs E1 and E2.

\citet{fer97} have measured the velocity of the H$\alpha$-emitting gas within a radius of $\sim$10\farc\ of the center of Per~A, approching to within 2\farcs6 of its nucleus.  Along the inner filament, the H$\alpha$ emission has a velocity of about $-25 {\rm \ km \ s^{-1}}$ at a radius of 2\farcs6 and about $-105 {\rm \ km \ s^{-1}}$ at a radius of 3\farcs5.  This radial range coincides with the outer portion of the inner filament where the CO(2-1) emission is primarily blueshifted: the H$\alpha$ velocity compares well with the range of CO(2-1) velocities seen here, including the trend of increasingly blueshifted velocities towards larger radii.  Along the eastern filament, the H$\alpha$ emission has a velocity of about $-165 {\rm \ km \ s^{-1}}$ at a radius of 5\farcs7, which coincides quite closely to the inner portion of the eastern filament.  Here, the CO(2-1) velocity reaches about $-180 {\rm \ km \ s^{-1}}$, again comparable with that seen in H$\alpha$.

\citet{con01} have measured the velocity of the H$\alpha$-emitting gas out to a radius of nearly 10\farc\ on the eastern side and nearly 20\farc\ on the western side, but with poorer spatial sampling.  At and close to the nucleus, the H$\alpha$ emission is redshifted, like in in our CO(2-1) map.  Further west at the location of the western filament, the gas is blueshifted, again as is seen in our CO(2-1) map.  We therefore conclude that the CO(2-1) features are not just spatially coincident with bright H$\alpha$ features but, where the velocity of the latter has been measured, also have comparable velocities.

\subsection{Correspondence with X-ray features}
In Figure~6, we overlay (black) contours of the integrated CO(2-1) intensity on a X-ray image of the center of the Perseus cluster taken from \citet{fab06}.  We also plot in (red) contours the radio emission of Per~A at 90~cm, produced by relativistic jets from the central AGN, taken from \citet{ped90} \citep[see also][]{fab02}.  The X-ray image shows cavities to the north and south of the nucleus that coincide with the radio jets.  These cavities are therefore believed to be giant bubbles blown by the radio jets into the surrounding X-ray gas.  The bulk of the molecular gas --- specifically, all the features located east and west of center --- lies (in projection) between the X-ray cavities.  The isolated blob N1 lies close (in projection) to a local X-ray brightening projected against the northern X-ray cavity, and the blob S1 near the inner rim of the southern X-ray cavity.

In Figure~7, we overlay contours of the integrated CO(2-1) intensity on a map of the lowest temperature (0.5~keV, or $\sim$$5 \times 10^6 {\rm \ K}$) X-ray gas detected by Chandra towards the center of the Perseus cluster, taken from \citet{fab06}.  The bulk of this relatively cool X-ray gas lies in an east-west band (curving north) across the center of the galaxy, located between the two X-ray cavities.  As remarked by \citet{fab06}, the bulk of this relatively cool X-ray gas closely resembles the spatial distribution of the H$\alpha$ gas \citep[see also Fig.~12 of][]{fab06}, although the mass of the X-ray gas at this temperature alone is an order of magnitude larger than that of the H$\alpha$ gas.  The bulk of the molecular gas detected in our image also has a spatial distribution that closely matches that of this relatively cool X-ray gas.

\section{Gas captured from a gas-rich galaxy?}
The origin of the molecular gas in Per~A has long been a subject of debate and controversy.  Arguments have been made for mergers with or cannibalisms of gas-rich galaxies, ram-pressure stripping of gas-rich galaxies passing close to or through Per~A, and a X-ray cooling flow.  Here, we examine whether the observed properties of the molecular gas are consistent with capture from gas-rich galaxies.

\subsection{Interactions with a foreground infalling galaxy}
Suggestions that the molecular gas in Per~A is captured from either ram-pressure stripping or mergers/cannibalisms of gas-rich galaxies are bolstered by evidence that Per~A may currently be interacting with a gas-rich galaxy.  The latter has been detected through its emission in H$\alpha$ \citep{min57,rub77,ken79,bor90,ung90,cau92,fer97}, as well as absorption in both atomic hydrogen (HI) gas \citep{dey73,eke76,has82,van83} and Lyman~$\alpha$ \citep{bri82} against the continuum emission of Per~A at a velocity around $8200 {\rm \ km \ s^{-1}}$.  Despite having a velocity that is $\sim$$3000 {\rm \ km \ s^{-1}}$ higher than that of Per~A, and hence commonly referred to as the high-velocity system (HVS), the detection of line absorption in both HI and Lyman~$\alpha$, as well as continuum absorption in X-rays \citep{fab00,gil04}, implies that the HVS lies in front of Per~A.

Evidence that this galaxy is not just a chance foregound galaxy but in the close vicinity of Per~A comes from the detection of H$\alpha$ emission at velocities intermediate between those of Per~A and the HVS \citep{fer97}.  This intermediate-velocity gas may comprise gas from the HVS entrained by the hot intracluster X-ray gas.  Furthermore, the ionized gas in the HVS appears to exhibit a complex velocity field \citep[e.g.,][]{bor90,ung90}, which together with the active star formation in this galaxy \citep{cau92} suggests that the HVS has been disturbed by the intracluster medium or interactions with Per~A.

It is highly unlikely, however, that the molecular gas in Per~A was captured from the HVS.  First, the velocity of the molecular gas is $\sim$$3000 {\rm \ km \ s^{-1}}$ lower than the velocity of the HVS, and is predominantly blueshifted with respect to the systemic velocity of Per~A rather than redshifted like the HVS.  Second, the HVS has not been detected in HI emission nor CO, placing a 3$\sigma$ upper mass limit of $2 \times 10^9 {\rm \ M_{\odot}}$ in atomic hydrogen gas \citep{van83} and $5 \times 10^9 {\rm \ M_{\odot}}$ in molecular hydrogen gas \citep{jaf87}.  The latter is about a factor of four lower than the mass of molecular hydrogen gas in Per~A.  Consistent with these arguments, \citet{gil04} place a lower limit of $\sim$60~kpc on the separation between the HVS and Perseus~A, and argue that any collision between the two galaxies still lies in the future.

\subsection{Previous encounters with gas-rich galaxies}
If captured, the molecular gas in Per~A more likely originated from previous encounters with other gas-rich galaxies.  Gas captured in this manner should exhibit a transverse (i.e., perpendicular to radial) as well as any radial spatial-kinematic components following the path of the interacting galaxy through or decaying orbit of the cannibalized galaxy in the cD galaxy.  In both cases, one might expect the captured gas to orbit and hence exhibit a global rotation about the cD galaxy.

Where imaged and well resolved, the CO gas in elliptical galaxies has until now been invariably found to be distributed in rotating disks \citep[e.g., see][and references therein]{you05}, suggesting that their gas was indeed captured from gas-rich galaxies.  Where only CO spectra (at good signal-to-noise ratio) are available, many exhibit double-horned line profiles indicative of rotating disks \citep[e.g.,][]{maz93,wik95,eva99,lim00,lim04}.  In Per~A, however, we detect no transverse spatial-kinematic components nor any ordered global motion indicative of rotation.  Instead, the most prominent features that we detected in Per~A --- the three filaments described in $\S3$ --- exhibit only radial spatial-kinematic components.  If each filament is attributed to capture from an individual gas-rich galaxy passing through Per~A, this would require three separate events with each gas-rich galaxy taking a projected path that crosses the center of Per~A.  Although this could be the future fate of the HVS, there is no independent evidence that three such highly unlikely events have occurred in the relatively recent past.

\section{Gas accreted from a X-ray cooling flow?}
The excellent spatial correspondence between the molecular gas and the relatively cool X-ray gas (Fig.~7), but not the warmer X-ray gas \citep[Fig. 12 of][]{fab06}, in Per~A provides circumstantial evidence that the molecular gas originates from a cooling flow.  In the simplest possible situation imaginable, gas accreted from such a cooling flow should simply flow radially inwards along the gravitational potential of the cD galaxy.  The gas should therefore exhibit only radial spatial-kinematic components, as is indeed observed for the three filaments.  Such a signature alone, however, is not a unique indicator of radial inflows.  In principle, gas undergoing radial outflows also can exhibit the same spatial-kinematic pattern.  We now examine in greater detail the properties of these filaments to see if we can distinguish between inflows or outflows.

\subsection{Correspondence with dust}
\subsubsection{Outer molecular filaments and blobs}
The eastern and western filaments, as well as the other blobs of molecular gas, exhibit only blueshifted velocities.  This implies inflow if the filaments and blobs are located on the far (back) side of Per~A, or outflows if located on the near (front) side.  To deduce which is the case, we have compared the spatial distributions of these filaments and blobs to optical images showing dust \citep[e.g.,][]{car98,kee01,con01} seen in silhouette against starlight from Per~A.  If these filaments and blobs are located on the near side of the galaxy, they ought to show a good correspondence with the silhouette dust.  On the contrary, if these filaments and blobs are located on the far side of the galaxy and at a large inclination to the plane of the sky, they may not show up strongly if at all in silhouette dust.  Examples of silhouette dust seen in elliptical galaxies with the HST-WFPC2 can be found in \citet{mar99}, \citet{ver99}, \citet{dek00}, and \citet{tra01}.  Inferred dust masses as small as 10--$100 {\rm \ M_{\odot}}$ are detectable, which for a ratio in mass of molecular hydrogen gas to dust of about 100 (as in our Galaxy) corresponds to molecular gas masses as small as just $10^3$--$10^4 {\rm \ M_{\odot}}$.
 
In Figure~8, we overlay the integrated CO(2-1) intensity on the HST-WFPC2 image of Per~A taken from \citet{car98}.  An especially prominent and complicated network of dust can be seen in silhouette towards the north of the nucleus, extending both east and even further west.  Much if not all of this dust likely belongs to the foreground infalling galaxy \citep[e.g.,][]{kee01,con01}.  First, the overall spatial distribution of this dust network closely resembles that of the H$\alpha$ gas in the HVS, but not that in Per~A itself \citep[e.g., compare the silhouette dust in Fig.~7 with the individual H$\alpha$ images of the HVS and Per~A in Fig.~3 of][]{cau92}.  In addition, the morphology of this dust closely matches that of gas detected in X-ray absorption against Per~A, attributed to the HVS \citep{gil04}.

Whether belonging to the HVS or Per~A, or portions in both, neither the eastern nor western filaments, nor the blobs E1, E2, W1, and N1 coincide with any of the silhouette dust (S1 is located beyond the region spanned by the HST-WFPC2 image).  Specifically, the complicated network of dust does not extend to the location of the eastern filament and blobs, where the light distribution appears to be smooth.  Dust in silhouette is visible to the north, but not at the location, of the western filament and W1.  The light distribution at and immediately around the western filament and W1 appears to be quite smooth, although its proximity to the prominent silhouette dust does make any weaker silhouette features difficult to discern.  The lack of correspondence with silhouette dust suggests that both the eastern and western filaments, as well as the other blobs of molecular gas, are located on the far side of the galaxy and at a relatively large inclination to the plane of the sky.  Their blueshifted velocities therefore imply inflow.  We disregard the possibility that these filaments and blobs lack dust because they formed from metal-poor gas; on the contrary, the iron abundance of the X-ray gas is at least $\sim$0.5~solar \citep{sch02,chu02} within $\sim$100~kpc.

\subsubsection{Inner molecular filament}
To permit a more detailed comparison between the inner molecular filament and the silhouette dust, in Figure~9 we overlay the integrated CO(2-1) intensity on just the inner portion of HST-WFPC2 image of Per~A taken from \citet{kee01}.  The latter is a transmission map, representing the residual intensities after subtracting smooth model profiles for both the foreground and background galaxy \citep[for more details, see][]{kee01}.   Brighter regions therefore indicate intensities above the smooth background, and darker regions intensities below the smooth background (i.e., dust extinction).

Confirming what was seen in Figure~8, neither the western molecular filament nor blob W1 obviously coincide with the silhouette dust.  There is a patch of dust at the innermost radii of the western filament where the CO(2-1) emission is strongest, but this dust patch does not trace the length of the filament and hence is likely coincidental.  There is patchy filamentary dust at and around the location of the inner molecular filament, although once again this dust does not obviously trace out the filament.  If the dust seen here also belongs to the HVS, this would make any silhouette dust close to the nucleus of Per~A difficult to discern.  Thus, at the present time, we cannot be certain whether the inner filament is associated with any silhouette dust, and hence whether this filament is located on the near or far side of Per~A.

\subsection{Correspondence with radio jets}
The relative locations of the molecular features and radio jets from Per~A were shown earlier in Figure~6.  If the molecular filaments we observe correspond to gas entrained by the radio jets, they should lie in projection against, or at the edges of, these jets.  In addition, the filaments should have their long axes aligned approximately north-south, along the axis of the radio jets.  Instead, the molecular filaments have their long axes aligned approximately east-west, closely orthogonal to the axis of the radio jets.  Furthermore, all the filaments lie neatly between the X-ray cavities, in regions where there is presumably no energetic particles to excavate the X-ray gas.  Thus, the molecular filaments are unlikely to comprise gas entrained by the radio jets or other energetic particles.  The same argument applies to the blobs E1, E2, and W1.

On the other hand, the blobs N1 and S1 do lie in projection against the radio jets.  S1 is spatially resolved, and appears to be elongated in an approximately west-west direction, closely parallel with the western filaments.  If entrained by the radio jet, one would expect this blob to be elongated approximately north-south, along the axis of the radio jet.

\subsection{Free-fall in gravitational potential of Per~A}
Both the eastern and western filaments exhibit a linear increase in blueshifted velocities with decreasing radii.  Here, we examine whether such a motion can be produced by gas in free-fall along the gravitational potential of Per~A.

\subsubsection{Model for gravitational potential}
To model the gravitational potential of Per~A, we use an analytical model proposed by \citet{her90} that has been found to reproduce the projected surface brightness profile of elliptical galaxies parameterised only by their mass and effective radius.  In this scheme, the mass density as a function of radius is given by
\\
\begin{equation}
\rho(r) = {M \over 2 \pi} {a \over r} {1 \over (r + a)^3} \ ,
\end{equation}
\\
where $M$ is the total mass of the galaxy, $r$ the radius, and $a$ a scale length.  The scale length $a$ is related to the half-mass radius, $r_{1/2}$, which contains half of the mass of the galaxy by
\\
\begin{equation}
r_{1/2} = (1 + \sqrt2)a \ .
\end{equation}
\\
The gravitational potential is then given by
\\
\begin{equation}
\phi(r) = - {G M \over {r+a}} \ ,
\end{equation}
\\
where $G$ is the gravitational constant.  The velocity, $v(r)$, of an object in free-fall can be computed from the relation
\\
\newcommand{\ud}{\mathrm{d}}
\begin{equation}
{v(r)^2 \over 2} - {GM \over (r + a)} = {v(r_o)^2 \over 2} - {GM \over (r_o +a)} \ ,
\end{equation}
\\
where $r_o$ is the initial radius at which the object is released and $v(r_o)$ its initial velocity.

Taking values from \citet{smi90} but scaled to the Cosmology used here, Per~A has an optical V-band luminosity of $1.1 \times 10^{11} {\rm \ L_{\odot}}$ and an estimated mass-to-light ratio in this band of $M/L \sim 7.4$.  The inferred mass of the galaxy is therefore $8.3 \times 10^{11} {\rm \ M_{\odot}}$; given the uncertainty in $M/L$ (which, moreover, may change with radius), there is probably considerable uncertainty in the estimated galaxy mass.  \citet{smi90} quote an effective radius, $R_e$, for Per~A of 12.4~kpc.  The effective radius is related to the scale length $a$ by
\\
\begin{equation}
R_e \approx 1.8153 a \ ,
\end{equation}
\\
and so $a \approx 6.8 {\rm \ kpc}$.

\subsubsection{Predicted velocity profile}
\subsubsubsection{Outer filaments}
The X-ray gas is very nearly in hydrostatic equilibrium, and so a given parcel of cooling gas is initialy at rest with respect to the cluster.  In Figure~10, we plot from Eq.~(4) how an object, which is initially at rest, changes in velocity with radius when dropped from two different initial radii, at 8.5~kpc (upper panel) and 12.2~kpc (lower panel).  (The choice of initial radii is explained in the next paragraph.)  For comparative purposes, but also because of the considerable uncertainty in the estimated galaxy mass, we show the results for two different masses for Per~A of $3.4 \times 10^{11} {\rm \ M_{\odot}}$ (solid curve) and the value estimated above of $8.3 \times 10^{11} {\rm \ M_{\odot}}$ (dashed curve).  As can be seen, beyond a distance of $\sim$1~kpc from which the object is dropped, the velocity increases essentially linearly with decreasing radius, as is observed for both the eastern and western filaments, irrespective of the assumed galaxy mass.  The challenge here is to reproduce not just the observed velocity gradient, but also the actual velocities spanned by the two filaments.

With a suitable choice for the inclination of the filaments as the only free parameter, we find that this simple model can reproduce the measured range of velocities of both the eastern and western filaments provided we adopt the lower value for the galaxy mass of $3.4 \times 10^{11} {\rm \ M_{\odot}}$.  In Figure~10, we overplot the PV-diagrams of the western and eastern filaments shown in Figure~4 but deprojected in both velocity and radius for inclinations that give the best model fit.  As discussed in $\S5.1.1$, these two filaments must be located at a relatively large inclination to the plane of the sky (and on the far side of Per~A) for dust in silhouette not to be seen associated with these filaments.  We find relatively good fits to the predicted curves for an inclination of about $40^{\circ}$ for the western filament and $47^{\circ}$ for the eastern filament.  Assuming larger inclinations would result in a larger velocity gradient consistent with a higher galaxy mass (dashed curves in Fig.~9), but lower deprojected velocities placing these filaments on the part of the curve where the velocity increases non-linearly with decreasing radius.  Assuming smaller inclinations, on the other hand, would result in a smaller velocity gradient inconsistent with the model (unless one is prepared to accept an even lower galaxy mass).  The choice of initial radii mentioned in the previous paragraph is therefore determined by our model once a galaxy mass is adopted and the best fit inclinations for the filaments determined.

For an inclination of $40^{\circ}$, the western filament (upper panel) spans deprojected radii of 4.2--7.4~kpc.  In this model, the outermost tip of the filament lies just 1.1~kpc away from the location where it first began its free fall in the gravitational potential of Per~A.  The deprojected velocity of this filament at its outermost radius is about $-125 {\rm \ km \ s^{-1}}$ compared with the computed model value of about $-121 {\rm \ km \ s^{-1}}$, and at its innermost radius about $-280 {\rm \ km \ s^{-1}}$ compared with the computed model value of about $-273 {\rm \ km \ s^{-1}}$.  For an inclination of $47^{\circ}$, the eastern filament (lower panel) spans deprojected radii of 8.1--11.3~kpc, with its outermost tip lying just 0.9~kpc away from where it began its free fall (i.e., at a radius of 12.2~kpc).  The deprojected velocity of this filament at its outermost radius is about $-55 {\rm \ km \ s^{-1}}$ compared with the computed model value of about $-90 {\rm \ km \ s^{-1}}$, and at its innermost radius about $-190 {\rm \ km \ s^{-1}}$ compared with the computed model value of about $-206 {\rm \ km \ s^{-1}}$.  We regard the agreement between this simple model and our observations to be satisfactory given that we have ignored a number of effects, primarily the outward pressure the X-ray gas exerts on the infalling molecular gas.  The latter should cause an infalling object to increase more slowly in velocity with decreasing radius (i.e., resulting in a shallower velocity gradient) than predicted by our simple model, thus requiring a larger galaxy mass to produce the observed velocity gradient than assumed in our best-fit model.  This would bring our results into better agreement with model predictions for the inferred larger galaxy mass for Per~A of $8.3 \times 10^{11} {\rm \ M_{\odot}}$.

The scenario we have in mind then is the following.  The CO(2-1) transition is excited at a critical density in molecular hydrogen gas of $\sim$$10^3 {\rm \ cm^{-3}}$.  From an analysis of the line ratios in CO(2-1) and CO(1-0) based on single-dish observations, \citet{bri98} also infer a density for the molecular hydrogen gas of $\sim$$10^3 {\rm \ cm^{-3}}$, and a temperature of 20--30~K.  By comparison, the density of the X-ray gas at radii of 5--10~kpc from the center of Per~A lies in the range 0.05--$0.1 {\rm \ cm^{-3}}$ \citep{chu02,fab06}.  The western filament corresponds to molecular gas that has condensed from X-ray gas located at a distance of $\sim$8.5~kpc from the center of Per~A.  The cooling gas decouples from the X-ray gas, and then free falls in the gravitational potential of Per~A.  By the time it has travelled 1.1~kpc from its initial location, taking about 18.2~Myr, the cooling infalling gas has already been (largely) converted to molecular gas and has attained a density sufficiently high and temperature sufficiently low to be detectable in CO(2-1).  Because the filament is relatively dense, it is unlikely to be (significantly) disrupted during its passage through the more diffuse X-ray gas.  Furthermore, its linear appearence suggests that the filament is not significantly perturbed by turbulence or any other disturbances.  The same picture applies for the eastern filament, except that in this case the molecular gas condensed from X-ray gas initially located at a distance of $\sim$12.2~kpc from the center of Per~A.  The time it takes for the cooling infalling gas to travel 0.9~kpc from its initial location to the outermost tip of the eastern filament is about 21.1~Myr.

\subsubsubsection{Inner filament}
The measured properties of the inner filament, on the other hand, are difficult to explain with a simple model.  Consider the situation where this filament also corresponds to radially-infalling molecular gas, as we have shown is likely the case for the two outer filaments.  The blueshifted portion of the filament would then have to lie on the far side of the galaxy, and the redshifted portion on the near side.  As mentioned in $\S3$, these two portions of the filament may actually be separate features that are poorly resolved spatially in our map.  Unlike the two outer filaments, however, the blueshifted portion of the inner filament exhibit smaller blueshifted velocities at smaller radii (see top panel of Fig.~4).  It may be that, close to the center of the galaxy, the infalling molecular gas is being retarded by the high pressure of the X-ray gas, such that the blueshifted portion of the filament experiences an inward decrease in velocity.  The redshifted portion of the filament, on the other hand, exhibits larger redshifted velocities at smaller radii, implying an inward increase in velocity.

Alternatively, and perhaps more plausibly, this filament may correspond to infalling molecular gas that is settling into the gravitational potential well of Per~A.  Specifically, the gas may be settling into orbits to form a dynamically unrelaxed disk around the nucleus.  \citet{hec85} tentatively infer an approximately constant stellar orbital velocity of $\sim$$50 {\rm \ km \ s^{-1}}$ over radii of $\sim$5\farc--25\farc\ in Per~A.  It is reasonable to expect that as one approaches the center of Per~A, the X-ray gas increasingly partakes in the overall galaxy rotation.  In such a case the molecular gas deposited by the X-ray cooling flow should also possess angular momentum, and cannot directly flow into the center but must eventually form a rotationally-supported structure (i.e., disk) around the nucleus.

In support of this picture, hot ($\sim$1000~K) molecular hydrogen gas has been detected in the near-IR vibrational line of H$_2$~$\nu = 1-0$~S(1) around the nucleus of Per~A \citep{wil05}.  This gas appears to be distributed in a disk-like structure oriented east-west with a radius of $\sim$50~pc, and therefore shares the same morphological orientation as the inner filament.  The eastern side of the H$_2$~$\nu = 1-0$~S(1) disk is redshifted in velocity by $\sim$$100 {\rm \ km \ s^{-1}}$ and the western side blueshifted by $\sim$$100 {\rm \ km \ s^{-1}}$, therefore spanning the same range of velocities and also having the same sense in velocity gradient as the inner filament.  The inner filament may therefore trace the outer, cooler, dynamically unrelaxed portion of the inner, warmer, and more dynamically relaxed molecular hydrogen disk, and is responsible for despositing gas into this circumnuclear molecular disk.

\section{Implications for Cooling Flow}
\subsection{Mass-deposition rate}

\subsubsection{Outer molecular filaments}
Given that the eastern and western filaments exhibit a linear dependence in velocity with radius, we can compute their dynamical ages from the relationship
\\
\begin{equation}
T_{dyn} = {r_{max} - r_{min} \over v(r_{max}) - v(r_{min})} \ \ ln\left({v_{max} \over v_{min}}\right)  \ ,
\end{equation}
\\
where $v(r_{max})$ is the deprojected velocity at the maximum deprojected radial extent $r_{max}$, and $v(r_{min})$ is the deprojected velocity at the minimum deprojected radial extent $r_{min}$.  For the values inferred in $\S5.3.2$, we derive a dynamical age of $\sim$16~Myr for the western filament and $\sim$29~Myr for the eastern filament.  The analytical model described in Sect 5.3.2 (where the predicted velocity changes close to but not perfectly linearly with radius) gives a dynamical age of 21.7~Myr for the western filament and 16.5~Myr for the eastern filament.  Changing the inferred inclinations of the filaments by $\pm 20^{\circ}$ only changes their dynamical age by a factor of $\pm 2$ (the dynamical age is proportional to tan($i$), where $i$ is the inclination of the filaments to the plane of the sky).

Given that the western filament has a molecular gas mass of $\sim$$8.5 \times 10^8 {\rm \ M_{\odot}}$ and the eastern filament $\sim$$6.1 \times 10^8 {\rm \ M_{\odot}}$ as listed in Table~1, the total mass-deposition rate into these two filaments is therefore $\sim$$75 {\rm \ M_{\odot} \ yr^{-1}}$.  This value should be considered as a very rough estimate that is probably accurate to only a factor of a few.  It would be instructive to compare the inferred mass-deposition rate into these two molecular filaments with the upper limit in mass-deposition rate inferred for the X-ray gas at the same radius.  At the present time, however, no such information for the Perseus Cluster has been published.  Instead, below we compare the inferred mass-deposition rate with that inferred for other gas components below X-ray temperatures.

\subsubsection{Other gas components}
\citet{bre06} have detected O~VI lines tracing ionized gas at temperatures of $\sim$$10^{5.5} {\rm \ K}$ in a 30\farc\ square region centered on Per~A.  They attribute these lines to gas from a X-ray cooling flow, and infer a mass-deposition rate of $32 \pm 5 {\rm \ M_\odot \ yr^{-1}}$ within a radius of $\sim$6~kpc.  The inferred mass-deposition rate is roughly comparable with that inferred for the band of X-ray gas at $\sim$$5 \times 10^6 {\rm \ K}$ located between the X-ray cavities, provided this gas originates from a cooling flow \citep{fab06}.  If responsible in its entirety for producing (at the present mass-deposition rate) the inner molecular filament, which has a mass of $1.8 \times 10^9 {\rm \ M_\odot}$ (Table~1), the cooling flow in the inner $\sim$6~kpc region of the cluster must have lasted at least $\sim$56~Myr.  The mass-deposition rate inferred by \citet{bre06} is, coincidentally, roughly comparable with the mass-deposition rate we infer for the two outer molecular filaments.  These filaments, however, lie beyond the region spanned by the observed O~VI lines.  In addition, there is molecular gas lying even further out than the region we have mapped (and which also may originate from a cooling flow), suggesting that the X-ray cooling flow in the Perseus cluster may have a significantly higher mass-deposition rate than that inferred from the O~VI lines alone.

The close spatial-kinematic correspondence between the cool molecular gas detected here and ionized gas at $\sim$$10^{4} {\rm \ K}$ traced in H$\alpha$ suggests that at least a portion of the latter could also originate from a X-ray cooling flow.  The mass in H$\alpha$ gas over its entire visible extent is only $\sim$$10^7 {\rm \ M_{\odot}}$ \citep{hec89}, two orders of magnitude below the mass in molecular gas of the filaments reported here, and three orders of magnitude below the total mass of molecular gas in Per~A.  The bulk of the gas from the X-ray cooling flow must therefore spend relatively little time at temperatures of $\sim$$10^{4} {\rm \ K}$.

One of the outstanding challenges is to understand the time evolution of the cooling flow as it transforms from X-ray-emitting plama into molecular gas.  At a radius of $\sim$10~kpc where we infer the two outer molecular filaments to first cool from the X-ray gas, the azimuthally-averaged cooling time of the X-ray gas is $\sim$500~Myr \citep{dun06}.  Although the cooling flow must spend relatively little time at temperatures of $\sim$$10^4 {\rm \ K}$, it may well spend much longer times at other temperatures before condensing into molecular hydrogen gas.  Specifically, the time it spends at temperatures above those traced in CO(2-1) may be considerably longer than the inferred free-fall time of just $\sim$20~Myr to reach the outer tips of the molecular filaments.  If the cooling gas remains coupled to the X-ray gas by magnetic fields, and strong magnetic fields have been inferred in the central region of Per~A \citep{tay06}, then this gas may not start to free-fall until atomic hydrogen has (almost fully) recombined at temperature below $\sim$4000~K.

\subsection{AGN fueling and feedback}
Molecular gas accreted from a X-ray cooling flow is an excellent candidate for fueling the AGN observed in many cD galaxies lying in rich clusters.  In the case of Per~A, the circumnuclear hot molecular hydrogen disk detected by \citet{wil05} may be responsible for fueling the AGN; this disk in turn may have been accreted from or is replenished by the inner cool molecular filament observed here (\S5.3.2).  The powerful radio jets produced by the AGN reheats the surrounding X-ray gas, thereby diminishing if not quenching the cooling flow \citep[e.g.,][]{chu02,bir04}.  If the cooling flow is severely disrupted, the AGN may eventually run out of fuel and turn off.  This allows the cooling flow to eventually resume and deposit a fresh supply of fuel to the AGN.

In the absence of any disturbances (i.e., reheating of the X-ray gas), one would expect the gas deposited by a X-ray cooling flow to stream inwards along all radial directions.  Instead, all three molecular filaments detected in Per~A are confined to the east-west directions, reflecting the global distribution of the molecular gas in this galaxy \citep[][]{sal06}.  This is just the configuration expected if the cooling flow is quenched (or reduced below the level of detectability) over a large solid angle in the north-south direction by the radio jets from Per~A, but is still able to operate in directions orthogonal to these jets.  Indeed, as pointed out in $\S3.4$, the bulk of the molecular gas detected coincides with a band of relatively cool X-ray gas lying between the X-ray cavities inflated by the radio jets.  Our observations implicate this relatively cool X-ray gas as part of the cooling flow.

Note that, when deprojected in three dimensions, the molecular gas is not infalling along all directions orthogonal to the radio jets, but is instead confined to certain radial directions mostly toward the far side of the galaxy.  This suggests that, even orthogonal to the radio jets, large portions of the X-ray gas may be experiencing reheating, reducing their cooling flow if any to below the level of detectability.  Indeed, the X-ray image of the Perseus cluster shows ripples in virtually all directions projected in the plane of the sky, but which is strongest over a broad range of angles centered along the north-south directions \citep{fab06}.  Each ripple is believed to be an acoustic wave driven by episodic expansions in the X-ray bubbles inflated by the radio jets from Per~A.  Once a given bubble reaches a sufficiently high pressure to become buoyant it rises through the X-ray gas, whereby the radio jet creates new X-ray bubbles at the center.  The observed ripples (acoustic waves) are believed to gently heat the surrounding gas.  Our results indicate that a (detectable) cooling flow is still able to operate in a few localized regions where presumably the X-ray gas is not being strongly reheated.  Even at these locations, the cooling flow is likely intermittent or modulated in time by disturbances propagating through the X-ray gas.

\citet{dun06} infer an average duty cycle for the bubbling activity in putative cooling-flow clusters of at least $\sim$$70\%$, and perhaps as high as $\sim$$90\%$.  If the cooling flow is severely disrupted if not completely quenched when the AGN is on, such a high duty cycle may not permit the X-ray gas to both cool and reach the central supermassive black hole in the cD galaxy during the relatively short time when the AGN is off.  Instead, our results imply that a residual cooling flow occurs even when the AGN is on, and that this cooling flow is able to provide a nearly continuously supply of fuel --- most likely in the form of molecular gas --- to the AGN.

\section{Summary and Conclusions}


To better understand the origin of the molecular gas in Per~A, we have imaged this galaxy in CO(2-1) at an angular resolution of 3\farc\ (spatial resolution of 1~kpc) with the SubMillimeter Array (SMA).  Single-dish maps with angular resolutions as high as 12\farc\ (4~kpc) have previously shown that the molecular gas is concentrated in the east-west direction, and extends $\sim$40\farc\ ($\sim$14~kpc) in both directions from center.  This gas shows no discernible ordered spatial-kinematic pattern, exhibiting blueshifted velocities both east and west of center, and redshifted velocities at or close to the center.  Our SMA map (Figs.~1 and 2), which covers a region of radius $\sim$28\farc\ ($\sim$10~kpc) centered on Per~A, reveals that:

\begin{itemize}

\item[1.] the molecular gas extends to the edge of our field, and is concentrated primarily in three radial filaments with projected lengths ranging from at least 1.1--2.4~kpc spanning the galaxy center to 7.7~kpc.  These filaments, along with several blobs, have a total mass in molecular gas of $(4.2 \pm 0.2) \times 10^9 {\rm \ M_{\odot}}$, with the three filaments alone containing $\sim$$80\%$ of the detected gas mass.  

\item[2.] the bulk of the molecular gas detected (i.e., all three filaments and all except two blobs of molecular gas) is aligned roughly east-west and lie between two X-ray cavities to the north and south of center.  This gas is aligned approximately orthogonal to the axis of the radio jets from the AGN in Per~A.


\item[3.] the two outer (western and eastern) filaments, and the other blobs of molecular gas, exhibit only blueshifted velocities.  The velocities of both these filaments change linearly with radius to larger blueshifted velocities at smaller radii.  Neither these filaments nor the blobs are coincident with detectable silhouette dust, suggesting that they lie at a relatively large inclination to the plane of the sky on the far (back) side of Per~A.

\item[4.] the inner filament exhibits both blueshifted and redshifted velocities that straddle the systemic velocity of the galaxy.  Unlike the two outer filaments, the velocity of the inner filament does not change linearly with radius; furthermore, this filament exhibits smaller blueshifted or larger redshifted velocities at smaller radii.  It is not clear whether this filament is coincident with silhouette dust associated with Per~A.

\item[5.] all the molecular gas detected in the east-west direction is spatially coincident with H$\alpha$ gas, often lying against or close to local H$\alpha$ brightenings.  Where measured, specifically at the locations of the inner and western filaments, the velocities of the H$\alpha$ gas are comparable with that of the molecular gas.  The molecular gas also coincides spatially with a band of relatively cool X-ray gas lying between the two X-ray cavities.

\end{itemize}

We interpret the results in the following manner:

\begin{itemize}

\item[1.] the filaments are unlikely to have been captured from one or more gas-rich galaxies.  Gas captured in this manner should orbit and hence show a global rotation about the cD galaxy, as has hitherto been the case in all other elliptical galaxies that have been well mapped in CO.

\item[2.] lying on the far (back) side of Per~A, the two outer filaments are therefore flowing radially inwards towards the center of the galaxy.  The only known process that can naturally deposit gas in this manner is a X-ray cooling flow.  

\item[3.] the velocity pattern of the two outer filaments can be easily reproduced as free fall in the gravitational potential of Per~A, with their outer tips lying $\sim$1~kpc downstream from where the cooling gas first decouples from the X-ray gas.  Both filaments have a dynamical age of $\sim$20~Myrs, implying a total mass-deposition rate into these filaments of roughly $\sim$$75 {\rm \ M_{\odot} \ yr^{-1}}$ (likely accurate to only a factor of a few).

\item[4.] the velocity pattern of the inner filament cannot be explained with a simple model.  We speculate that this filament likely traces gas settling into the gravitational potential well of Per~A, and perhaps corresponds (in part) to the outer regions of a putative hot molecular hydrogen gas disk with a radius of $\sim$50~pc around the AGN.

\end{itemize}

Despite the strong and widespread heating of the surrounding gas inferred from X-ray observations, our results provide the most direct evidence yet for a (residual) X-ray cooling flow in the Perseus cluster.  This cooling flow may provide a nearly continuous supply of fuel for the AGN in Per~A, which in turn regulates the cooling flow.

\acknowledgments
We thank A. C. Fabian and J. S. Sanders for kindly providing a fits file of the X-ray image, and C. J. Conselice for a fits file of the H$\alpha$ image, used in this paper.  We thank the second of the two anonymous referees for suggesting changes that improved our presentation.  Y.-P. Ao acknowledges the ASIAA for supporting his stay as a Visiting Scholar, when much of this work was done.  J. Lim thanks Jan Vritlek and Bill Forman for useful discussions during a visit to the Harvard CfA, and acknowledges a grant from the National Science Council of Taiwan in support of this work.  The Submillimeter Array (SMA) is a collaborative project between the Smithsonian Astrophysical Observatory and the Academia Sinica Institute of Astronomy \& Astrophysics of Taiwan.

\clearpage

\begin{deluxetable}{cccccccc}
\tabletypesize{\scriptsize}
\tablecaption{Parameters of Molecular Features}
\tablewidth{0pt}
\tablehead{
\colhead{Name} & 
\colhead{Inner radius} & \colhead{Outer radius} & 
\colhead{Position angle} &
\colhead{Inner velocity} & \colhead{Outer velocity} &
\colhead{Flux density} & \colhead{Gass mass}
\\
\colhead{} & 
\colhead{(asec)} & \colhead{(asec)} & 
\colhead{(degrees)} &
\colhead{$(\rm km \ s^{-1})$} & \colhead{$(\rm km \ s^{-1})$} &
\colhead{$(\rm Jy \ km \ s^{-1})$} & \colhead{$(\rm \times 10^8 \ M_{\odot})$}
}
\startdata
Inner filament & $1.2 \pm 0.3$   & $4.2 \pm 0.3$   & $\sim$$270$ & $+100$ & $-100$ & $62.1 \pm 3.6$ & $18.0 \pm 1.1$   \\
Western filament & $9.0 \pm 0.3$ & $15.7 \pm 0.4$  & $\sim$$281$ & $-180$ & $-80$  & $29.3 \pm 2.6$ & $8.5 \pm 0.8$   \\
W1             & $21.6 \pm 0.6$  & \nodata         & $\sim$$198$ & $-20$  &\nodata & $9.9 \pm 1.1$  & $2.9 \pm 0.3$   \\
Eastern filament & $15.2 \pm 0.4$ & $21.3 \pm 0.3$ & $\sim$$76$  & $-160$ & $-40$ & $21.0 \pm 2.6$ & $6.1 \pm 0.8$   \\
E1             & $25.7 \pm 0.5$  & \nodata         & $\sim$$76$  & $-70$  &\nodata & 
$14.9 \pm 1.7$ & $4.3 \pm 0.5$   \\
E2             & $22.8 \pm 0.7$  & \nodata         & $\sim$$63$  & $-80$  &\nodata & $4.2 \pm 0.8$ & $1.2 \pm 0.2$   \\
N1             & $18.2 \pm 0.4$  & \nodata         & $\sim$$358$ & $-30$  &\nodata & $4.9 \pm 0.8$ & $1.4 \pm 0.2$   \\
S1             & $20.9 \pm 0.6$  & \nodata         & $\sim$$211$ & $-130$ &\nodata & $8.4 \pm 1.3$ & $2.4 \pm 0.4$   \\
\enddata
\tablecomments{The flux densities listed have been corrected for the primary beam response of the SMA.  The inner/outer radii and velocities listed for the filaments are derived from the channel maps.  For the features W1, E1, E2, N1, and S1, which are mostly spatially unresolved, we list only the radii and velocity at their centroids as derived from the moment maps.}
\end{deluxetable}

\clearpage

\clearpage
\begin{figure}
\includegraphics[width=15.5cm]{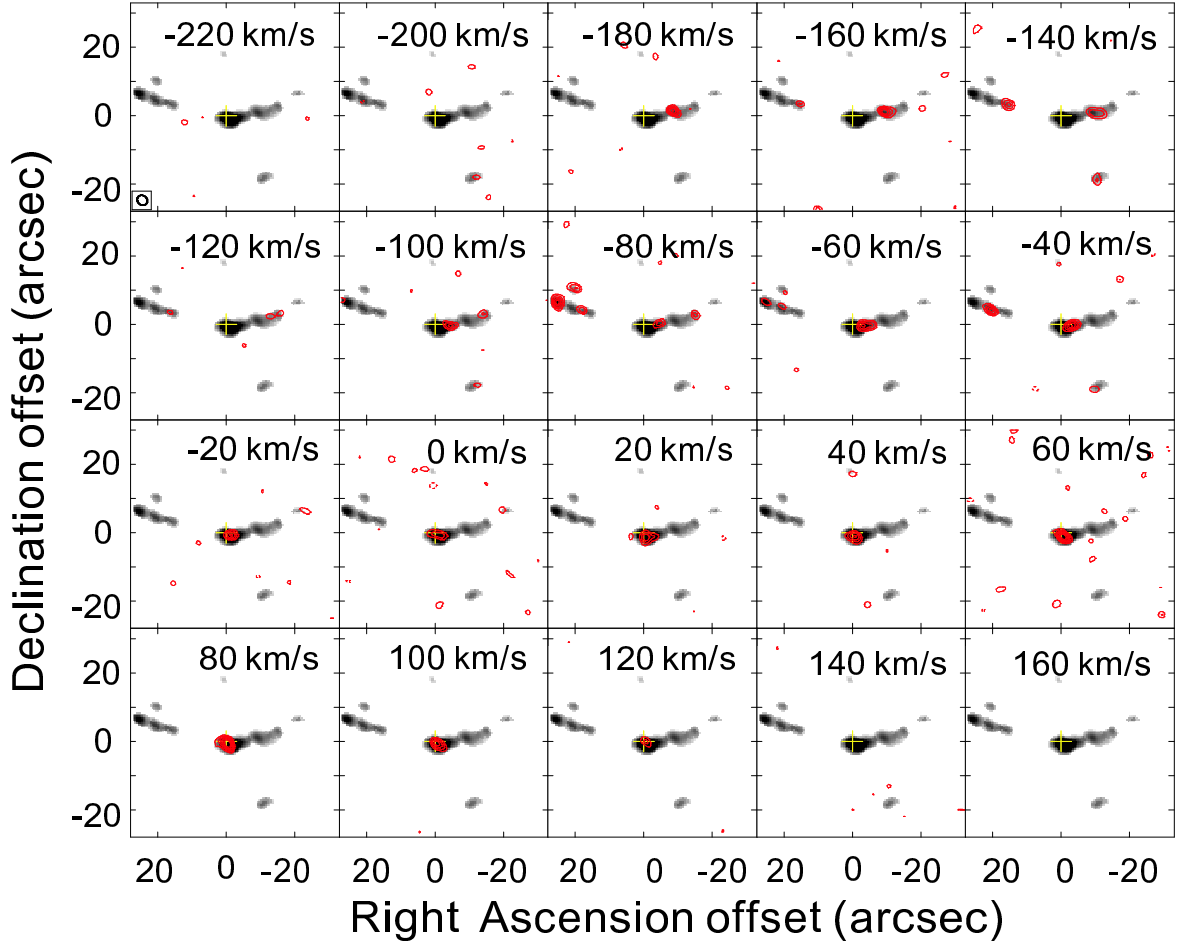}
\caption{Channel maps showing contours of CO(2-1) intensity with the velocity indicated in each panel measured with respect to the systemic heliocentric velocity of $5264 {\rm \ km \ s^{-1}}$ \citep{huc99}.  Contour levels are plotted at -3, 3, 4, 5, 6, 7, $8 \times 16 {\rm \ mJy \ beam^{-1}}$ (root-mean-square noise level; $1\sigma$).  No primary beam correction has been applied to these maps.  The gray background corresponds to the integrated CO(2-1) intensity after primary beam correction as shown in Figure~2.  The position of the nuclear continuum source in Per~A is indicated by a cross.  The synthesized beam is shown as an ellipse at the lower left corner of the top left panel, and has a size of $3\farcs0 \times 2\farcs7$ and position angle $39^{\circ}.5$.  The filaments indicated in Figure~2 are not uniformly bright in intensity along their lengths, and hence may not appear continuous in these channel maps.}
\end{figure}
\clearpage
\begin{figure}
\includegraphics[width=15.5cm]{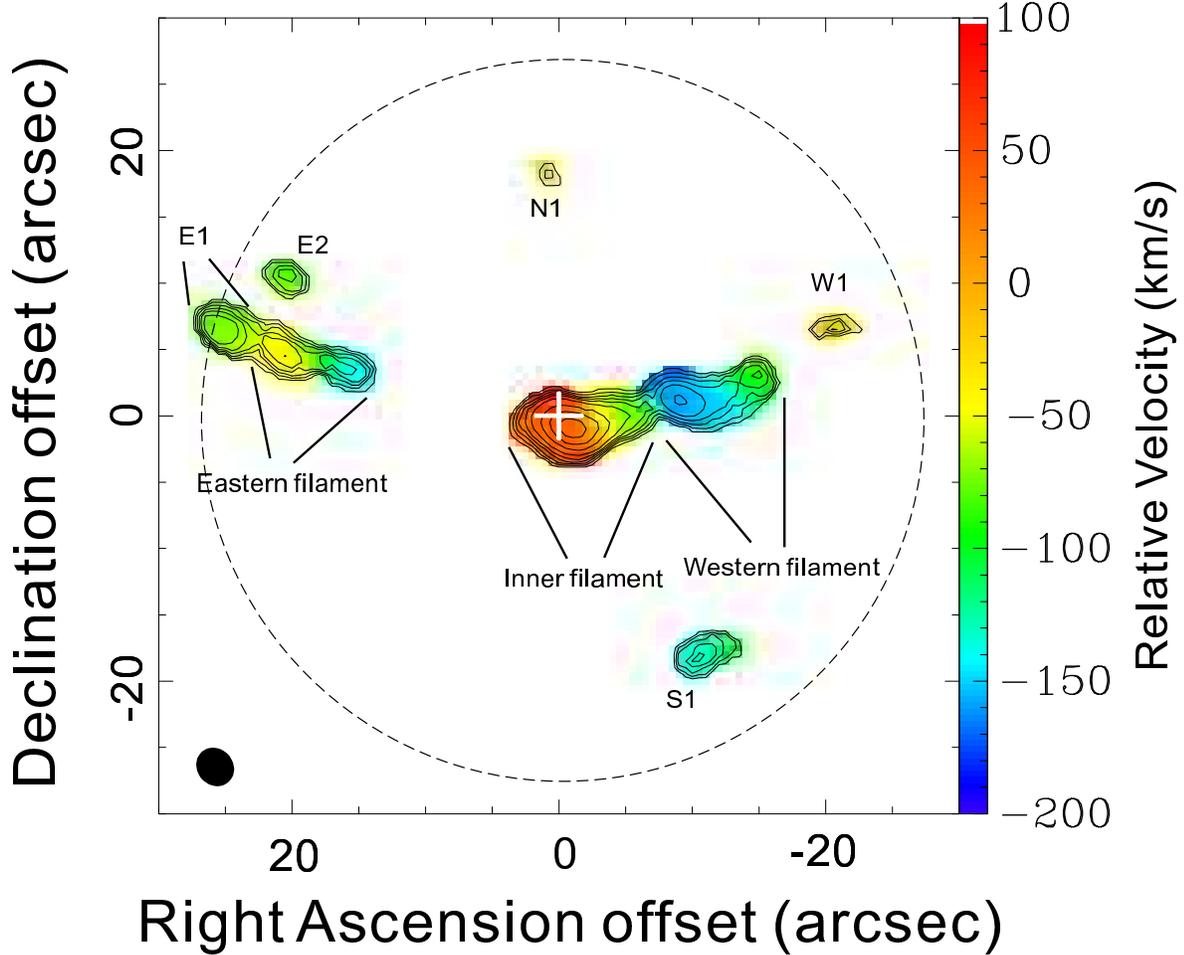}
\caption{Contours of the integrated CO(2-1) intensity plotted at levels of 3, 6, 9, 12, 15, 20, 30, 40, and $50 \times 0.32 {\rm \ Jy \ km \ s^{-1}}$, and color-coded map of the intensity-weighted CO(2-1) mean velocity measured with respect to the systemic heliocentric velocity of $5264 \pm 11 {\rm \ km \ s^{-1}}$ \citep[][]{huc99} as indicated by the right vertical bar.  The full-width half-maximum of the SMA primary beam is 55\farc\ (20~kpc), as indicated by the dotted circle.  The integrated CO(2-1) intensity has been corrected for the primary beam response.  The map is centered on the active nucleus of Per~A (${\rm Right \ Ascension = 03^h \ 19^m \ 48^s.16}$, ${\rm Declination}=41^{\circ} \ 30\farcm \ 42\farcs1$), which was detected in the continuum at the location indicated by a cross.  The synthesized beam is shown as a filled ellipse at the lower left corner, and has size of $3\farcs0 \times 2\farcs7$ with major axis at a position angle of $39^{\circ}.5$.  The inner, western, and eastern filaments, along with the features E1, E2, W1, N1, and S1, are labeled as described in the text.
}
\end{figure}
\clearpage
\begin{figure}
\includegraphics[width=.8\textwidth]{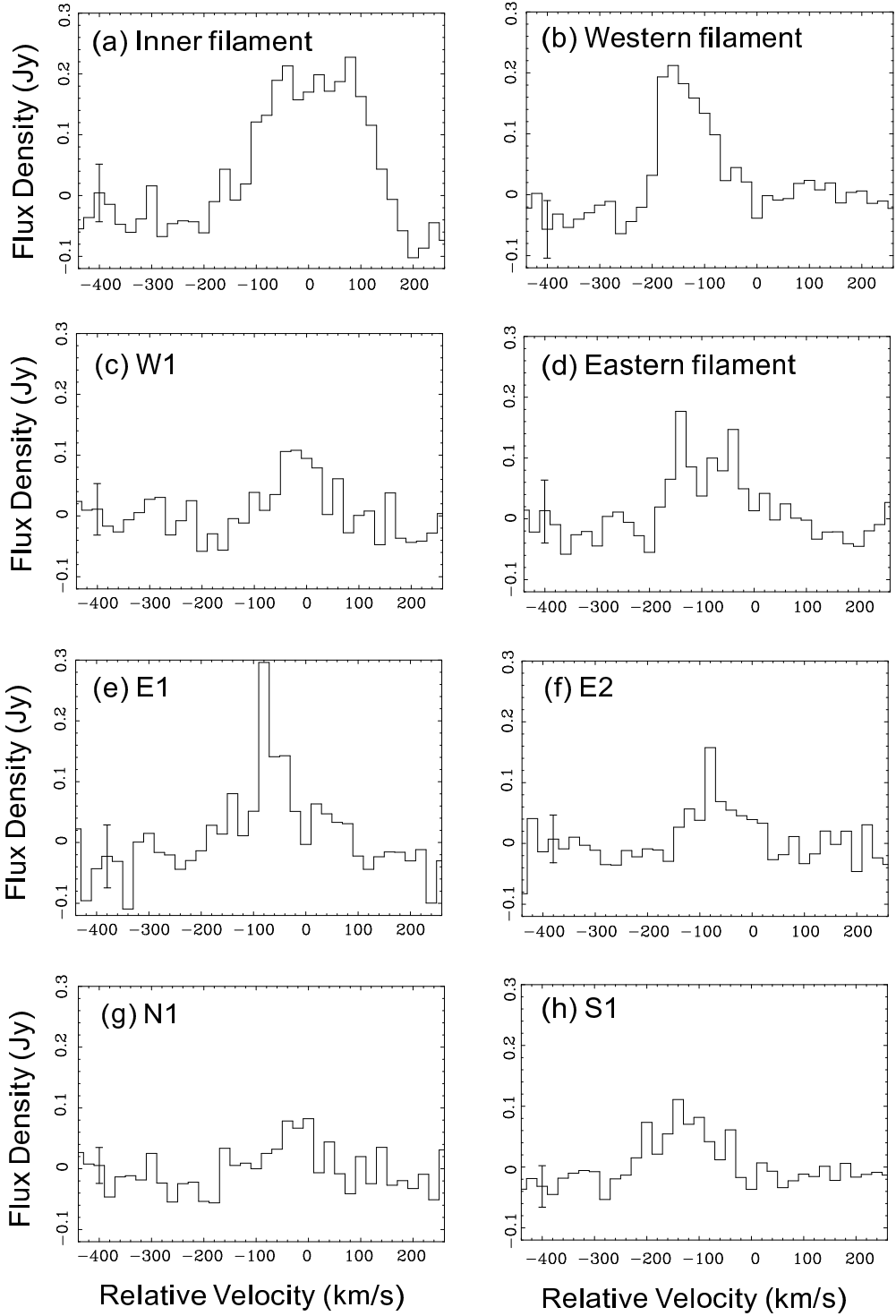}
\caption{Spectra of all the features labeled in Figure~2, derived from the channel maps shown in Figure~1 but after correcting for the primary beam response of the SMA.  An error bar corresponding to $\pm 1 \sigma$ uncertainty is plotted in each panel.
}
\end{figure}
\clearpage
\begin{figure}
\centerline{\includegraphics[width=.55\textwidth]{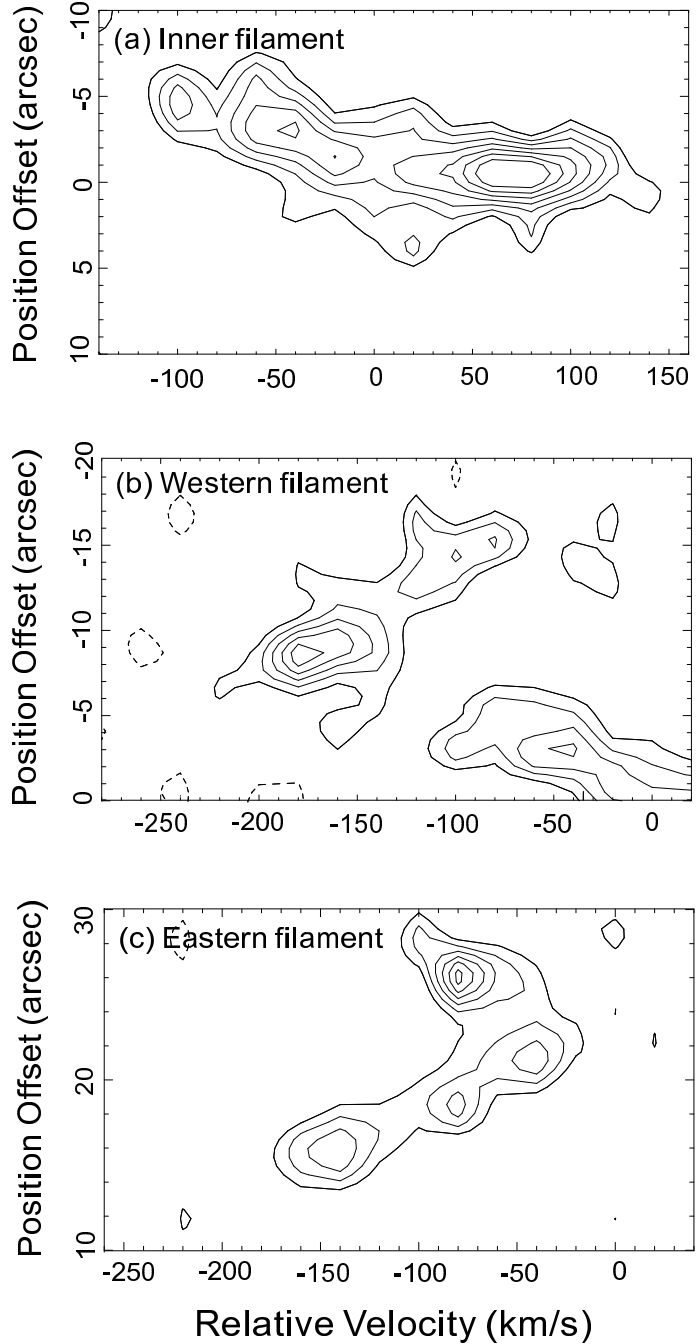}}
\caption{Position-velocity (PV) diagrams showing CO(2-1) intensity along the radial extent of (a) inner filament along a position angle of $270^{\circ}$, (b) western filament along a position angle of $281^{\circ}$, and (c) eastern filament along a position angle of $76^{\circ}$ as a function of velocity.  Contour levels are plotted at $-2$, 2, 3, 4, 5, 6, and $7 \times 16 {\rm \ mJy \ beam^{-1}}$ ($1\sigma$).  Positions correspond to offsets from the center of Per~A with negative values indicating west, and velocities are measured with respect to the systemic heliocentric velocity of $5264 \pm 11 {\rm \ km \ s^{-1}}$ \citep[][]{huc99}.  In panel (b), the feature at positions $\gtrsim -5\farc$ and velocities $\gtrsim -100 {\rm \ km \ s^{-1}}$ corresponds to the inner filament.  In panel (c), the feature at a position of 26\farc\ and velocity of $-80 {\rm \ km \ s^{-1}}$ that is discontinous in velocity from the eastern filament corresponds to E1 (Fig.~2), which is probably a separate structure as described in the text.
}
\end{figure}
\clearpage
\begin{figure}
\includegraphics[width=15.5cm,angle=-90]{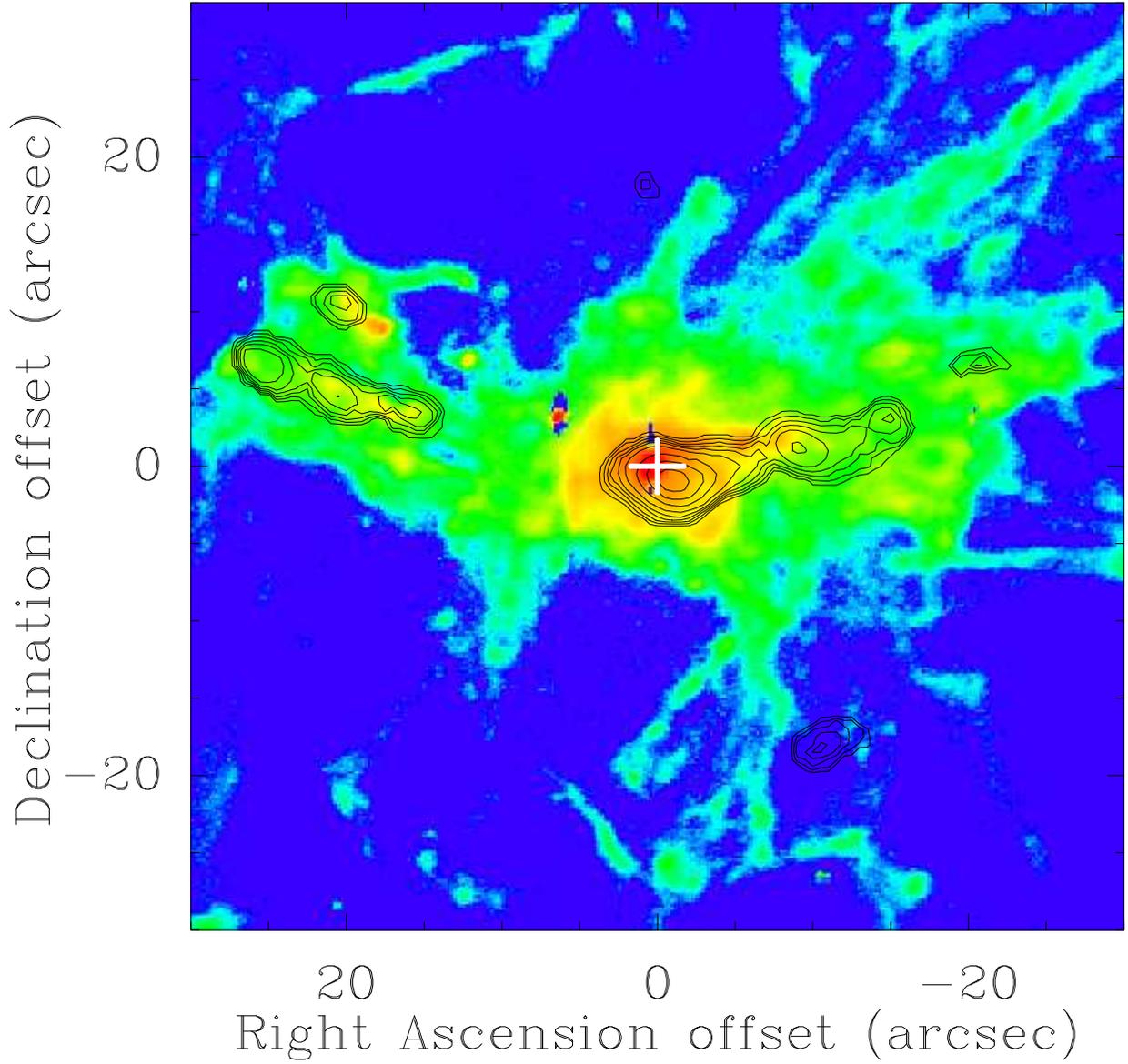}
\caption{Contours of the integrated CO(2-1) intensity (from Fig.~2) overlaid on a false-color map of the H$\alpha$ nebula associated with Per~A adapted from \citet{con01}.  This maps spans only the inner central region of the H$\alpha$ nebula.
}
\end{figure}
\clearpage
\begin{figure}
\includegraphics[width=15.5cm]{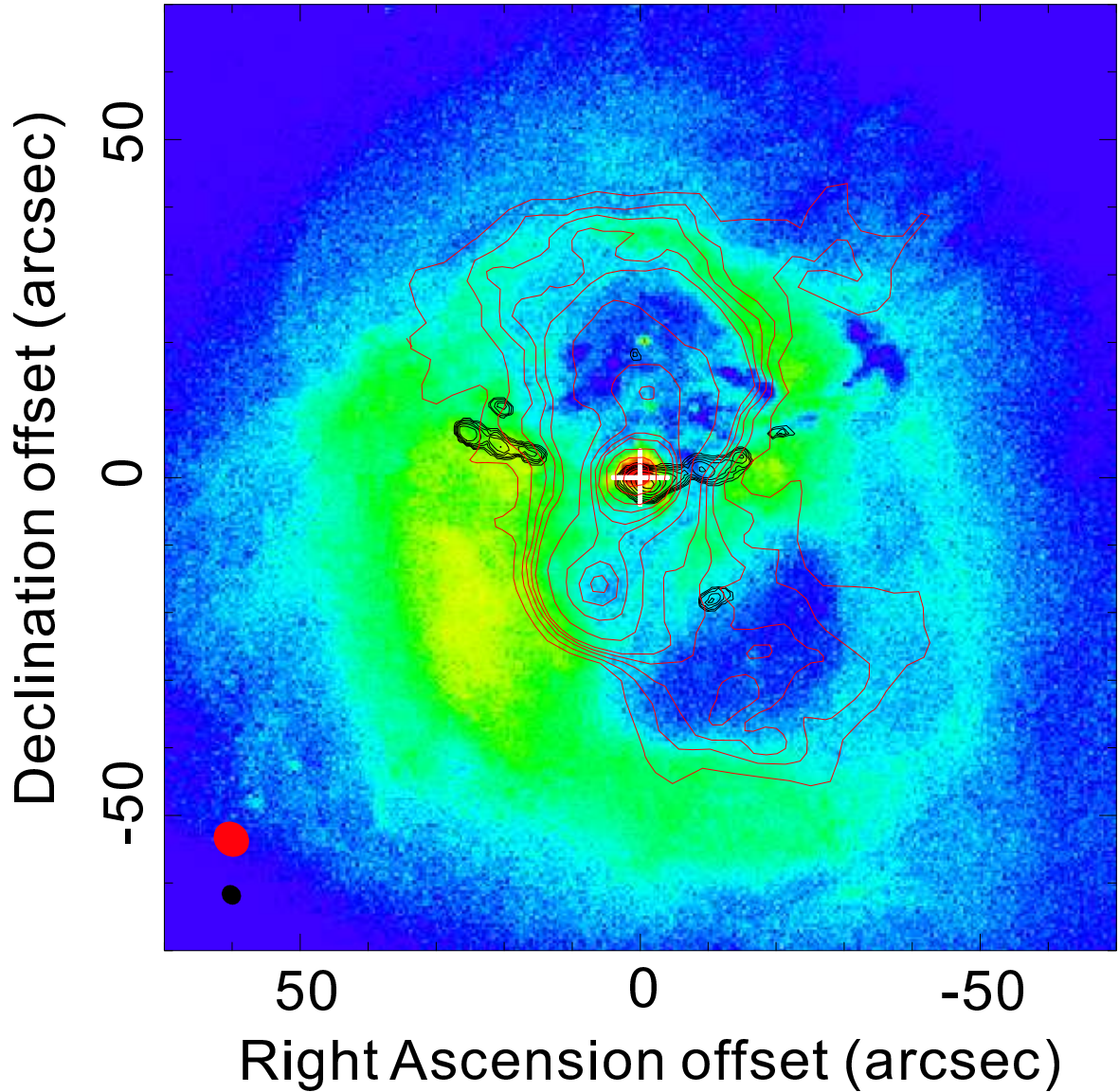}
\caption{Black contours of the integrated CO(2-1) intensity (from Fig.~2) and red contours of the continuum emission at 90~cm from \citet{ped90} showing radio jets from the AGN in Per~A, both overlaid on a false-color X-ray image from \citet{fab06}.  The synthesized beams of the CO(2-1) and 90~cm images are shown as the black and red ellipses, respectively, at the lower left corner.  
}
\end{figure}
\clearpage
\begin{figure}
\includegraphics[width=15.5cm,angle=-90]{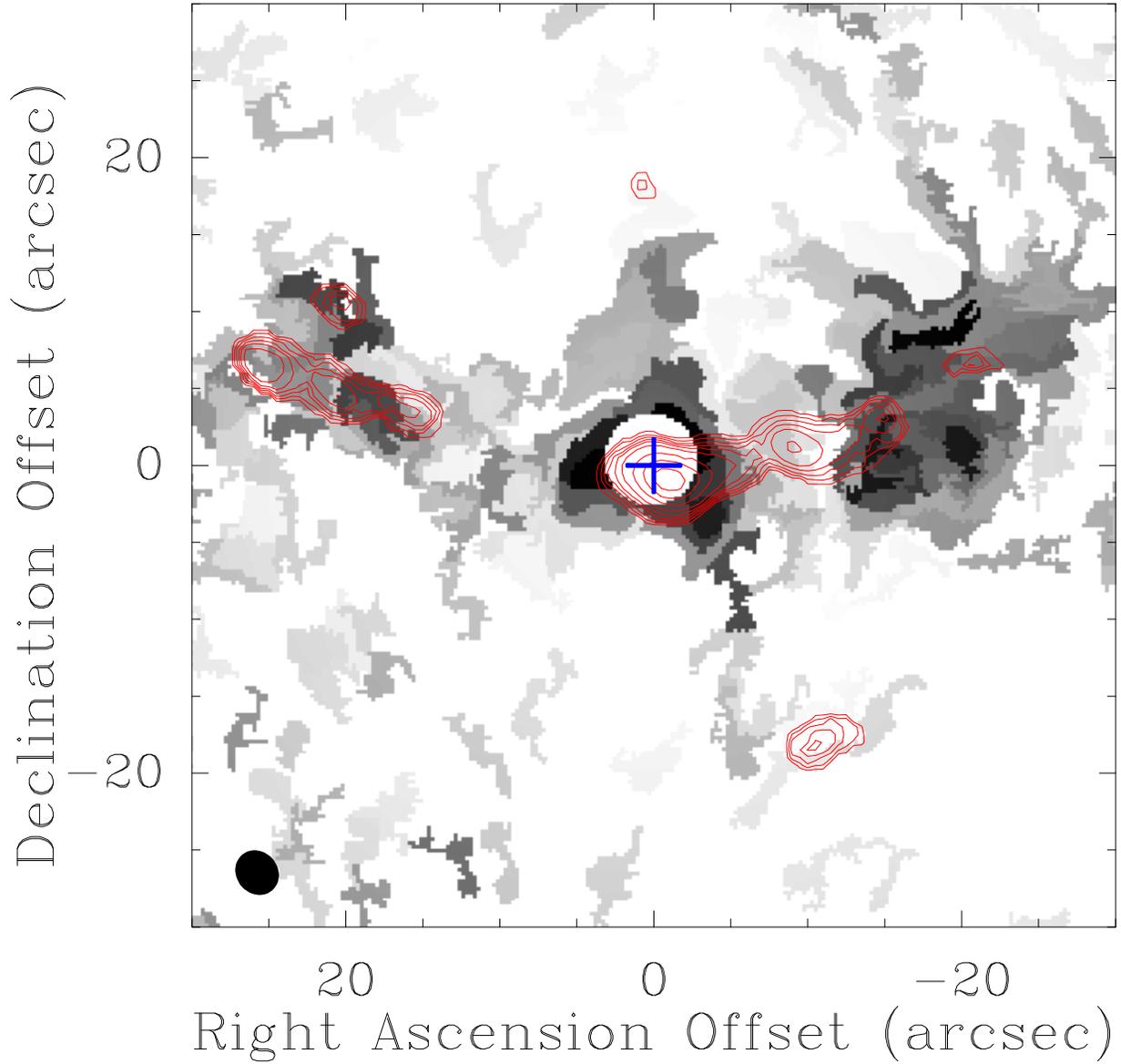}
\caption{Contours of the integrated CO(2-1) intensity (from Fig.~2) overlaid on a X-ray map of the inferred mass distribution of the 0.5~keV ($\sim$$5 \times 10^6 {\rm \ K}$) component in the core of the Perseus cluster \citep[from][]{fab06}.  
}
\end{figure}
\clearpage
\begin{figure}
\includegraphics[width=15.5cm]{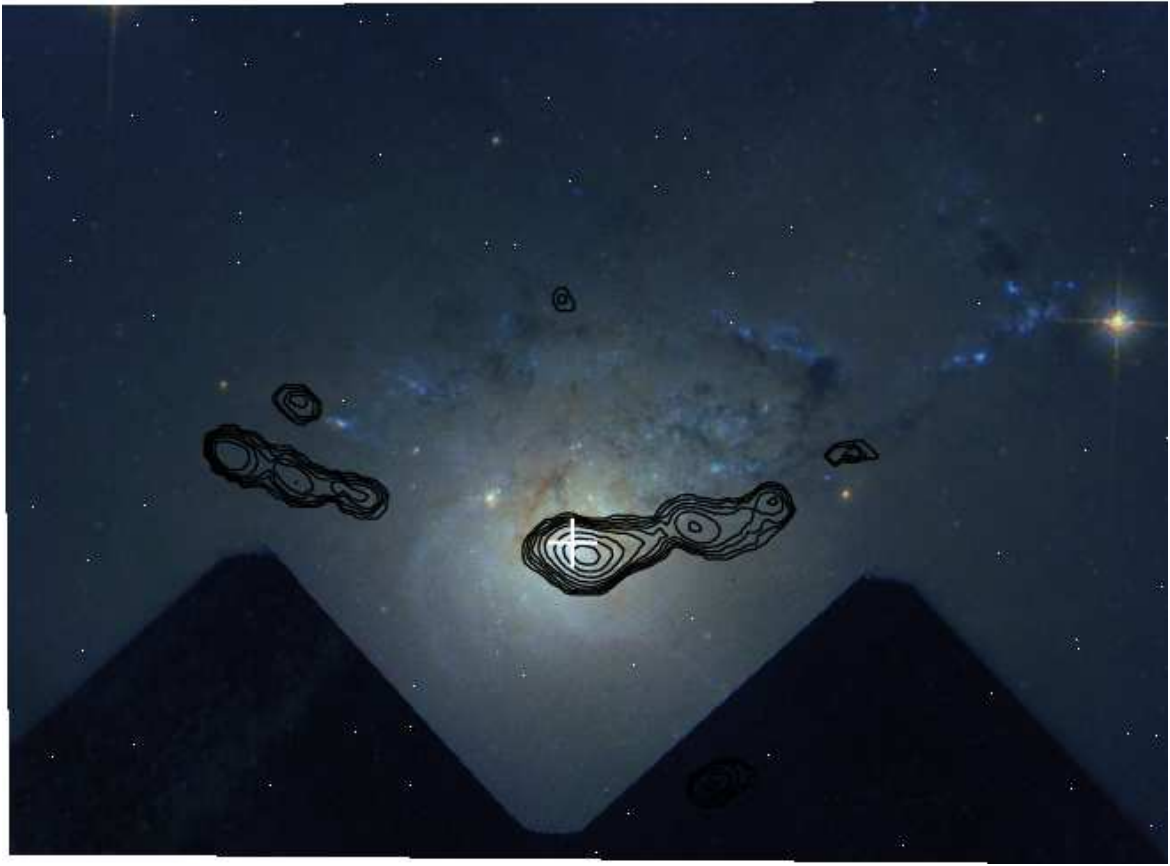}
\caption{Contours of the integrated CO(2-1) intensity (from Fig.~2) overlaid on the HST-WFPC2 optical image of Per~A from \citet{car98}.  Dark features are dust seen in silhouette against the optical continuum (primarily starlight) from Per~A, with a possible contribution also to the optical continuum from a foregound infalling galaxy (the HVS; see text).  The blob S1 lies outside the region spanned by this optical image.
}
\end{figure}
\clearpage
\begin{figure}
\includegraphics[width=15.5cm]{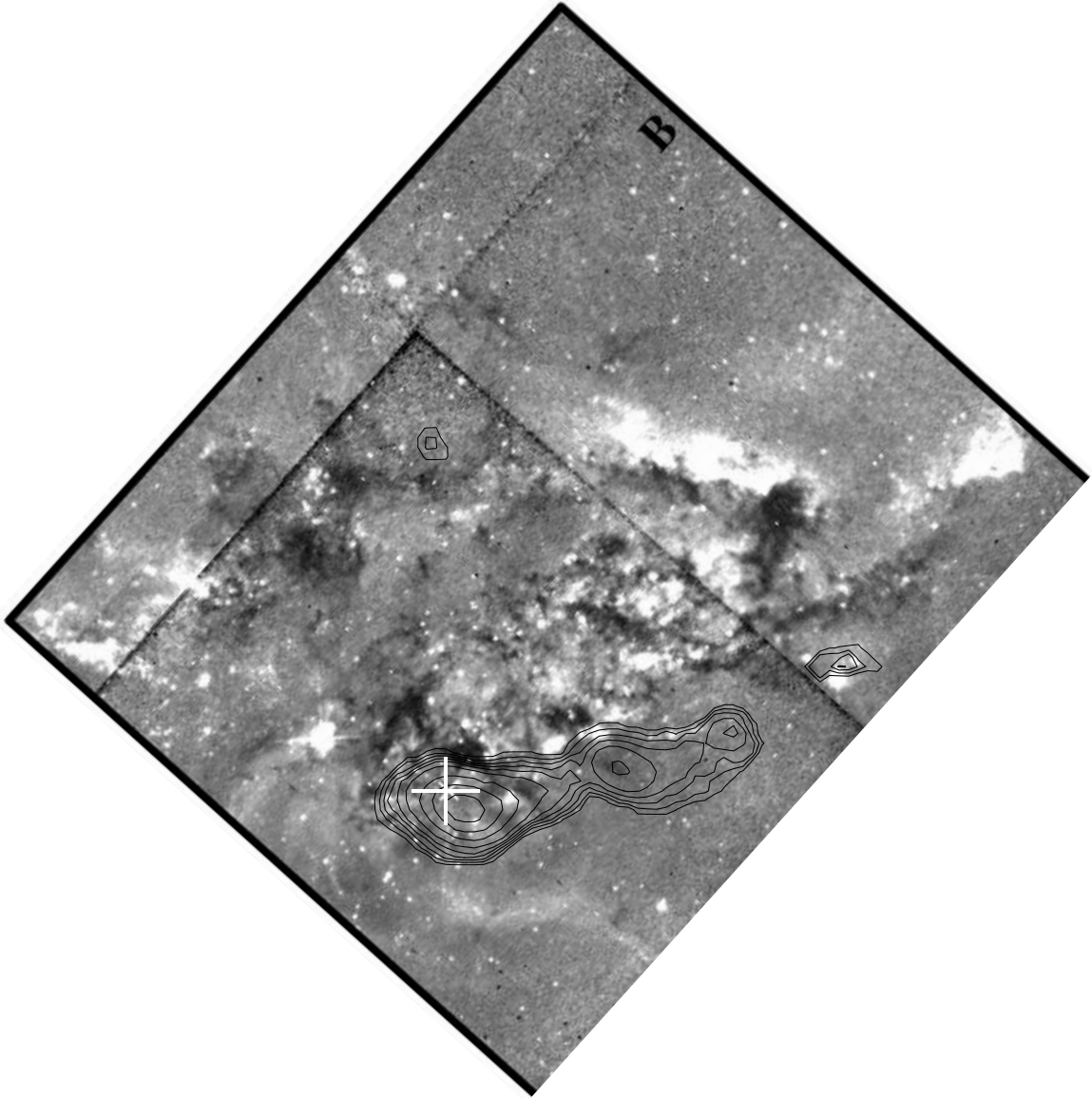}
\caption{Contours of the integrated CO(2-1) intensity (from Fig.~2) overlaid on the HST-WFPC2 B-band transmission map (see text) of the central region of Per~A from \citet{kee01}.  Bright features are intensities above a smooth background, and dark features below the same background.  Only the inner and western molecular filaments, along with the blobs W1 and N1, are visible over the region spanned by this optical map.
}
\end{figure}
\clearpage
\begin{figure}
\centerline{\includegraphics[width=.5\textwidth]{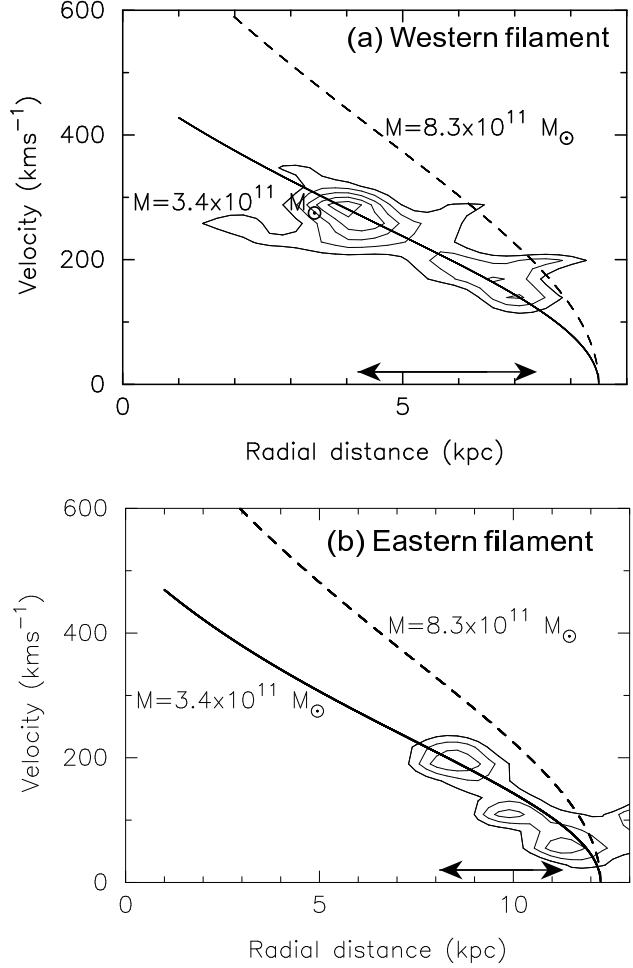}}
\caption{Velocity of an object in free-fall along the gravitational potential of Per~A as a function of radial distance from center, computed from Eq. (4) for two different galaxy masses as indicated by the solid and dashed curves.  Panel (a) shows the computed velocity profile of an object dropped from a radius of 8.5~kpc, and superposed the PV-diagram of the western filament (contours; from Fig.~4) deprojected by an inclination of $\sim$$40^{\circ}$ so that it spans 4.2--7.4~kpc as indicated by the double-headed arrow.  Panel (b) shows the computed velocity profile of an object dropped from a radius of 12.2~kpc, and superposed the PV-diagram (contours; from Fig.~4) of the eastern filament deprojected by an inclination of $\sim$$47^{\circ}$ so that it spans 8.1--11.3~kpc as indicated by the double-headed arrow.  As can be seen, the computed velocity profile assuming a lower galaxy mass of $3.4 \times 10^{11} {\rm \ M_{\odot}}$ (solid curve) better matches the observed PV-diagrams than the computed velocity profile for a galaxy mass of $8.3 \times 10^{11} {\rm \ M_{\odot}}$ inferred from the luminosity of the galaxy and its estimated mass-to-light ratio as described in the text.
}
\end{figure}

\end{document}